\documentclass[conference]{IEEEtran}
\usepackage{listings}
\def\BibTeX{{\rm B\kern-.05em{\sc i\kern-.025em b}\kern-.08em
    T\kern-.1667em\lower.7ex\hbox{E}\kern-.125emX}}

\usepackage[utf8]{inputenc}
\usepackage{xcolor}
\usepackage{color, colortbl}
\usepackage{graphicx,latexsym}
\usepackage{multicol}
\usepackage{multirow}
\usepackage{algorithm}
\usepackage{cite}
\usepackage{booktabs} % For formal tables
\usepackage{amsmath,amssymb,amsfonts}
\usepackage{threeparttable}
\usepackage{enumitem}
\usepackage{xurl}
\usepackage[square,sort&compress,numbers]{natbib}
\usepackage{color}
\usepackage{xspace}
\usepackage{amsmath}
\usepackage{graphicx}
\usepackage{subcaption}
\usepackage{caption}
\usepackage{algpseudocode}
\usepackage{adjustbox}
\usepackage{tikz}
\usepackage{soul}
\usepackage{mdwlist}
\usepackage{booktabs}
\usepackage{upquote}
\usepackage{multicol}
\usepackage{makecell, multirow}
\usepackage{easyReview}
\usepackage{pifont}
\usepackage{mdframed}
\usepackage[colorlinks]{hyperref}

\newcommand{\cmark}{\ding{51}}
\newcommand{\xmark}{\ding{55}}

\usepackage{array}
\newcommand{\etal}{\mbox{\emph{et al.\ }}}
\newcommand{\eg}{\mbox{\emph{e.g.,\ }}}
\newcommand{\ie}{\mbox{\emph{i.e.,\ }}}
\newcommand{\vs}{\mbox{\emph{v.s.\ }}}
\newcommand{\etc}{\mbox{\emph{etc}}}

\newcommand{\nucleus}{\textsc{Nucleus}\xspace}
\newcommand{\byteweight}{\textsc{ByteWeight}\xspace}

\newcommand{\angr}{\textsc{angr}\xspace}
\newcommand{\ghidra}{\textsc{Ghidra}\xspace}
\newcommand{\ida}{\textsc{IDA Pro}\xspace}
\newcommand{\dyninst}{\textsc{Dyninst}\xspace}
\newcommand{\ninja}{\textsc{Binary Ninja}\xspace}
\newcommand{\bap}{\textsc{BAP}\xspace}

\newcommand{\radare}{\textsc{Radare2}\xspace}

\newcommand{\jima}{\textsc{Jima}\xspace}

\newcommand{\ehf}{\texttt{eh\_frame}\xspace}

\newcommand{\tool}{FETCH\xspace}

\definecolor{dkgreen}{rgb}{0,0.6,0}
\definecolor{gray}{rgb}{0.5,0.5,0.5}
\definecolor{mauve}{rgb}{0.58,0,0.82}
\definecolor{codegreen}{rgb}{0,0.6,0}
\definecolor{codegray}{rgb}{0.5,0.5,0.5}
\definecolor{codepurple}{rgb}{0.58,0,0.82}
\definecolor{backcolour}{rgb}{0.95,0.95,0.92}

\floatstyle{ruled}
\newfloat{code fragment}{thp}{lop}
\floatname{code fragment}{Code Fragment}

\definecolor{dkgreen}{rgb}{0,0.6,0}
\definecolor{gray}{rgb}{0.5,0.5,0.5}
\definecolor{mauve}{rgb}{0.58,0,0.82}

\usepackage[scaled=.75]{beramono}
\usepackage[T1]{fontenc}

\newcommand{\binnum}{1,352\xspace}

\newcommand{\myparatight}[1]{\smallskip\noindent{\bf {#1}:}~}

\begin{document}
\title{Towards Optimal Use of Exception Handling Information for  Function Detection \\}

\author{              
        \IEEEauthorblockN{Chengbin Pang\IEEEauthorrefmark{1}\IEEEauthorrefmark{2}\textsuperscript{\textsection}\hspace{1.5mm}
                Ruotong Yu\IEEEauthorrefmark{2}\textsuperscript{\textsection}\hspace{1.5mm}
                Dongpeng Xu\IEEEauthorrefmark{3}\hspace{1.5mm}
                Eric Koskinen\IEEEauthorrefmark{2}\hspace{1.5mm}
                Georgios Portokalidis\IEEEauthorrefmark{2}\hspace{1.5mm}
                Jun Xu\IEEEauthorrefmark{2}
       }\\\vspace{-0.5em}
        \IEEEauthorblockA{
        \begin{tabular}{c c c}
                \IEEEauthorrefmark{1}Nanjing University  &
                \IEEEauthorrefmark{2}Stevens Institute of Technology   &
                \IEEEauthorrefmark{3}University of New Hampshire  \\
        \end{tabular}
        }
        \vspace{-2em}
}

\maketitle
\renewcommand{\footnoterule}
    {\noindent\smash{\rule[1pt]{0.49\textwidth}{0.4pt}}}
\begingroup\renewcommand\thefootnote{\textsection}
\footnotetext{These authors contributed equally to this paper. This work was done while Pang was a Visiting Scholar at Stevens Institute of Technology.}
\endgroup

\newcommand\added[1]{\textcolor{blue}{#1}}

\begin{abstract}
%Background%\jun{Done}

Function entry detection is critical for security of binary code. Conventional methods heavily rely on patterns, inevitably missing true functions and introducing errors. Recently, call frames have been used in exception-handling for function start detection. However, existing methods have two problems. First, they combine call frames with heuristic-based approaches, which often brings error and uncertain benefits. Second, they trust the fidelity of call frames, without handling the errors that are introduced by call frames.

In this paper, we first study the coverage and accuracy of existing approaches in detecting function starts using call frames. We found that recursive disassembly with call frames can maximize coverage, and using extra heuristic-based approaches does not improve coverage and actually hurts accuracy. Second, we unveil call-frame errors and develop the first approach to fix them, making their use more reliable.

\end{abstract}

\section{Introduction} 

%What is function detection and why it is hard
Function detection is the process of identifying code regions in binary software that are compiled from source-level functions. Accurate function detection is critical for guaranteeing the correctness and effectiveness of mainstream applications of binary security, ranging from binary code similarity detection~\cite{liu2018alphadiff,pewny2015cross}, legacy-code patching~\cite{wang2015reassembleable,wang2016uroboros,wang2017ramblr,bernat2011anywhere}, shadow stack protection~\cite{chen2015stackarmor}, coarse-grained~\cite{van2016tough,zhang2013control,erlingsson2006xfi,muntean2018tau} or fine-grained~\cite{wang2015binary,zhang2013practical,elsabagh2017strict,prakash2015vfguard,he2017no} control flow integrity, to code layout randomization~\cite{davi2015isomeron, wartell2012binary, koo2018compiler, hiser2012ilr,li2006address, williams2016shuffler, zhang2013practical,pappas2012smashing,koo2016juggling}.

%However, function detection has been widely regarded as being difficult.

The first step of function detection is to identify function entry points, or \emph{function starts} and this is, however,
% which is, however,
very challenging. First, symbols in a binary 
provide the true identity of function starts, but those symbols are normally stripped. Second, binary code is often riddled with complex constructs (\eg jump tables, tail calls, \etc) for performance optimization. %In binary code, source code level information such as function types and parameter list are missing. Furthermore, symbols in a binary -- the only source to unveiling function starts -- are normally stripped. 
Mainstream conventional approaches for function start detection~\cite{sok:sp20,meng2016binary,brumley2011bap} first recursively disassemble a given binary from known function starts (\eg program entry) and add the targets of call instructions as new function starts. They then scan the non-disassembled code to further detect function starts with common function prologues~\cite{radare2_org,shoshitaishvili2016sok} or data-mining models~\cite{brumley2011bap,meng2016binary}, followed by recursive disassembly again. Beyond such a hybrid approach, there are also solutions that  either (i) use data-mining models or neural networks to detect function starts~\cite{bao2014byteweight,shin2015recognizing}  or (ii) aggregate basic blocks connected by intra-procedural control flows into groups and consider the target of each call instruction or the first instruction in each group as a function start~\cite{andriesse2017compiler}. 

Although the above approaches have demonstrated some effectiveness in detecting function starts, they still share a fundamental drawback: they all attempt to recover function information using a pattern-driven principle, explicitly or implicitly.
This drawback impedes the adoption of those approaches in the context of security applications. In fact, the patterns used by them are usually 
% unreliable (when the source is unknown), 
incomplete (missing true function starts) and/or inaccurate (introducing false function starts).
% \textcolor{red}{EK: the first paren seems to be a different kind of thing than the second and third parens}
Unlike symbols whose reliability is guaranteed by compilers, the patterns collected by these approaches do not build on any reliable source. Inevitably, those approaches  lead to errors or omissions, which in turn reduces the confidence of users and even leads to cascading effects.

Recent advances~\cite{skochinsky2012compiler,williams2020egalito} have leveraged a new source to detect function starts in x64 binaries: call frames in the exception handling segment. To support exception handling, compilers emit call frames in x64 binaries as mandated by the ABI, giving information such as the start location for functions wherever possible. Mainstream binary analysis tools, in particular \ghidra~\cite{ghidra_org} and \angr~\cite{shoshitaishvili2016sok}, already use call frames to facilitate function start detection. However, we observe two common, critical problems. First, the tools try to improve coverage by mixing the use of call frames with additional approaches 
that are sometimes safe and sometimes unsafe.  Safe approaches leverage knowledge from the
binary (e.g., symbols), the machine (e.g., instruction set),
and/or the ABI (e.g., calling conventions) to provide correctness guarantees. However, unsafe approaches are also involved, which try to use common patterns but typically do not offer assurances of correctness. These unsafe approaches inevitably introduce errors, sabotaging the reliability of call frames and the safe approaches. Moreover, the benefits from unsafe approaches (\eg whether they can really improve coverage) remain unclear. Second, the tools fully trust the fidelity of call frames. They do not realize that call frames by themselves can also introduce errors and, not surprisingly, do not include any solutions to fix those errors. 

In this paper, we inspect the above two problems, aiming to bring new insights towards optimal strategies of using call frames for function start detection.

First, we study the coverage of existing tools, when combining call frames with other methods, and the accuracy of the results produced. To perform the study, we collected 1,395 binaries from both real-world application and the popular benchmarks, and we separately measured the coverage and accuracy of detecting function starts detected by each combination of approaches. Our key findings are (i) running safe recursive disassembly with call frames can already provide nearly full coverage; (ii) additionally running unsafe approaches from existing tools does not provide meaningful improvement to the coverage but, can introduce plenty of false positives.\footnote{False positive means the start of a function identified but it is actually not. False negative means the start of a function is not identified.} These bring insights towards both optimal coverage and better reliability in the use of call frames for function start detection.

%Second, we systematically unveil and quantify the errors that call frames can introduce. We further develop the first approach that can effectively fix nearly all the errors without introducing side effects. 

Second, we systematically unveil and quantify the errors that call frames can introduce. To be specific, we compared the function starts extracted from call frames and the ground truth in our benchmark binaries. We discovered that modern compilers keep separate call frames ({\em also separate symbols}) for distant parts in a non-contiguous function. When such call frames are directly used for function start detection, they can bring a significant group of false function starts. We also found that existing tools do not
provide any solution to handle such false function starts. Following our findings, we develop a new algorithm to fix errors brought by call frames. Our key insight is that distant parts in a non-contiguous function are typically connected via a jump. By checking that the jump between two call frames cannot be a jump between two functions (i.e, the jump cannot be a tail call), we can decide that the two call frames belong to the same non-contiguous function and thus, merge them. Inspired by this insight, we incorporate well-founded, restrictive criteria to detect tail calls, minimizing the chance of reporting false tail calls and ensuring that all missed tail calls are harmless. According to our evaluation, our algorithm can eliminate nearly 95\% of the false function starts introduced by call frames, without incurring harmful side effects. Further, all the missed false function starts are due to conservativeness of our implementation choices instead of the design of our algorithm.

% we adopt a completeness-driven but safe strategy to detect tail call.
% our approach essentially needs to examine the jump between two call frames and checks whether it is a tail call or not. In general, it is hard to design a perfect algorithm to detect tail calls. 

Our main contributions are as follows. 
\begin{itemize}[leftmargin=*]
\item {\em\bf New knowledge} -  We investigate the coverage and accuracy of function starts detected by combing call frames with different approaches from existing tools. We bring insights towards using call frames to achieve optimal coverage of function starts with a minimal hurt to the reliability.
\item {\em\bf New approach} - We are the first to systematically study, classify, and quantify the errors that call frames can bring. We develop the first approach that can fix the errors in call frames, making them a better information source for function start detection. 
\item {\em\bf New finding} - We unveil key problems in how existing tools use call frames and demonstrate their significance with quantitative evidence.
\item {\em\bf New tool} - We developed a tool incorporating all our strategies. Its source code is available at \url{https://github.com/ruotongyu/FETCH}.
\end{itemize}

%The rest of this paper is organized as follows. \S~\ref{sec:overview} gives an overview of our research problem, existing solutions, and our goals, followed by \S~\ref{sec:bcg-ehf} that demystifies exception handling at the binary level. \S~\ref{sec:study-cov} describes our study of the strategies towards optimal coverage and \S~\ref{sec:study-cov} presents our understanding about the errors due to call frames and our solution to fix the errors. \S~\ref{sec:discuss} discusses the limitations and future directions of our research. Finally, we conclude this paper in \S~\ref{sec:conclusion}.

\section{Overview}
\label{sec:overview}
% Move something to related work

% In this section, we first describe the function start detection in the context of binary code and then present the major challenges. Finally, we summarize the existing solutions as well as their remaining limitations and our goals.

\subsection{Problem Definition}

Informally, function detection is to reconstruct the mapping from the code in a binary to the corresponding functions in the source code. At the binary level, a function consists of a set of basic blocks, which has one entry point and one or more exit points. The principled solution of function detection is to find a function entry point firstly, or a \emph{function start}, and then follow the intra-procedural control flow to detect instructions until reaching the exit points. Accurately finding function starts is a universal foundation of function detection. In this paper, \emph{we, therefore, focus on function start detection}.

\subsection{Existing Solutions}

Past research has brought many solutions of function start detection. Most solutions developed in the earlier stage use three strategies. First, \byteweight~\cite{bao2014byteweight} and Shin \etal~\cite{shin2015recognizing} train decision trees and neural networks to detect function starts from raw binaries. Second, \nucleus~\cite{andriesse2017compiler} first recovers the instructions using linear sweep and then aggregates the instructions connected via intra-procedural control flows into groups. The target of a direct call instruction or the lowest address in each group is considered a function start. Third, the majority of tools (\eg \dyninst~\cite{meng2016binary}, \bap~\cite{brumley2011bap}, and \radare~\cite{radare2_org}) use a hybrid solution. The tools first gather symbols remaining in the binary and then run recursive disassembly from each symbol. The addresses of the symbols and targets of direct/indirect calls found in recursive disassembly are considered function starts. The tools finally detect function starts in the non-disassembled regions using common prologues or data mining models~\cite{sok:sp20}, followed by recursive disassembly from the newly discovered function starts.

A fundamental limitation shared by the above solutions is that, they all heavily rely on pattern matching or empirical learning to recover function starts from binary code. Even with the hybrid solutions, nearly 18\% of the function starts are detected by prologue matching (without counting the functions recursively found from those function starts)~\cite{sok:sp20}. The patterns and learned models can be incomplete or inaccurate and oftentimes over-fit the ``training'' data. As a consequence, those solutions inevitably introduce errors or miss true function starts. 

\myparatight{Using Exception Handling Information} Recently, many tools are adopting a more reliable source of information --- the exception-handling information --- to facilitate function start detection. Both \angr~\cite{shoshitaishvili2016sok} and \ghidra~\cite{ghidra_org} leverage call frames in the exception handling section to help detect function starts. They first consider the addresses recorded in existing symbols and call frames as function starts and run recursive disassembly from those addresses to detect more function starts carried by targets of call instructions. Then they take extra steps such as function prologue matching to find further missing function starts.  \jima~\cite{alves2019function} only leverages exception handling information to aid detection of exception handling code blocks.

The use of exception handling information by existing tools (\angr and \ghidra) in function start detection has two problems. First, they combine the reliable information in call frames with {\em unsafe approaches}, i.e., approaches that do
not offer assurances of correctness: \ding{182} both \angr and \ghidra run prologue matching to detect function starts in the non-disassembled code regions, followed by a round of recursive disassembly from each matched function start; \ding{183} both \angr and \ghidra leverage heuristics to detect tail calls and consider their targets as function starts ({\em not enabled by default}); \ding{184} \angr linearly scans the remaining code gaps and treats the beginning of each correctly disassembled code piece as a new function start~\cite{sok:sp20}. The use of unsafe approaches often bring errors, but it may not increase the coverage achieved by call frames and {\em safe approaches} that provide correctness guarantees (\eg recursive disassembly). Second, the existing tools fully trust the fidelity of call frames, without realizing and handling the errors that call frames can bring.

\begin{figure}[!t]
  \centering
\begin{lstlisting}[basicstyle={\scriptsize\ttfamily},numberstyle=\scriptsize,linewidth=0.9\columnwidth,xleftmargin=5mm]
double div(int a, int b) {
    if(b == 0) 
        throw "Division by zero error!";
    return (a/b);
}
void main (){
    int x = (int)getchar();
    int y = (int)getchar();
    try {
        return div(x,y);
    }
    catch (const char* msg){
        cerr << msg << endl;
    }
}
\end{lstlisting}
  \caption{An example of exception handling in C++ programs.}
  \label{fig:cppexc}
  \vspace{-0.8em}
\end{figure}

\subsection{Research Scopes}
In this paper, we focus on exploring the use of exception handling information for function-start detection. Our goal is not to develop a new approach from scratch. Instead, we aim to expose any shortcomings in how existing tools use exception handling information and identify the best strategies of using exception handling information for function start detection. Specifically, we have the following goals:

\begin{itemize}
    \item \textbf{Goal 1:} We study the coverage of function starts by combining call frames with both safe and unsafe approaches from existing tools. This will bring insights towards optimal coverage with a minimal threat to reliability. \S~\ref{sec:study-cov} discusses how we achieve this goal in detail. 
    \item \textbf{Goal 2:} We systematically study the errors that call frames can introduce and explore new solutions to fix the errors. This will help ensure the fidelity of call frames as an information source for function start detection. \S~\ref{sec:study-accuracy} presents our approach to the second goal.
\end{itemize}

In accordance to our goals, we restrict our discussion in this paper on binaries with call frames. To this regard, we focus on System-V x64 binaries (\eg x64 binaries running on Linux or other Unix variants) because the corresponding ABI~\cite{lu2018system} mandates the existence of call frames while the other types of binaries may not have call frames.

\begin{figure}[t!]
  \centering
  \includegraphics[width=0.95\linewidth]{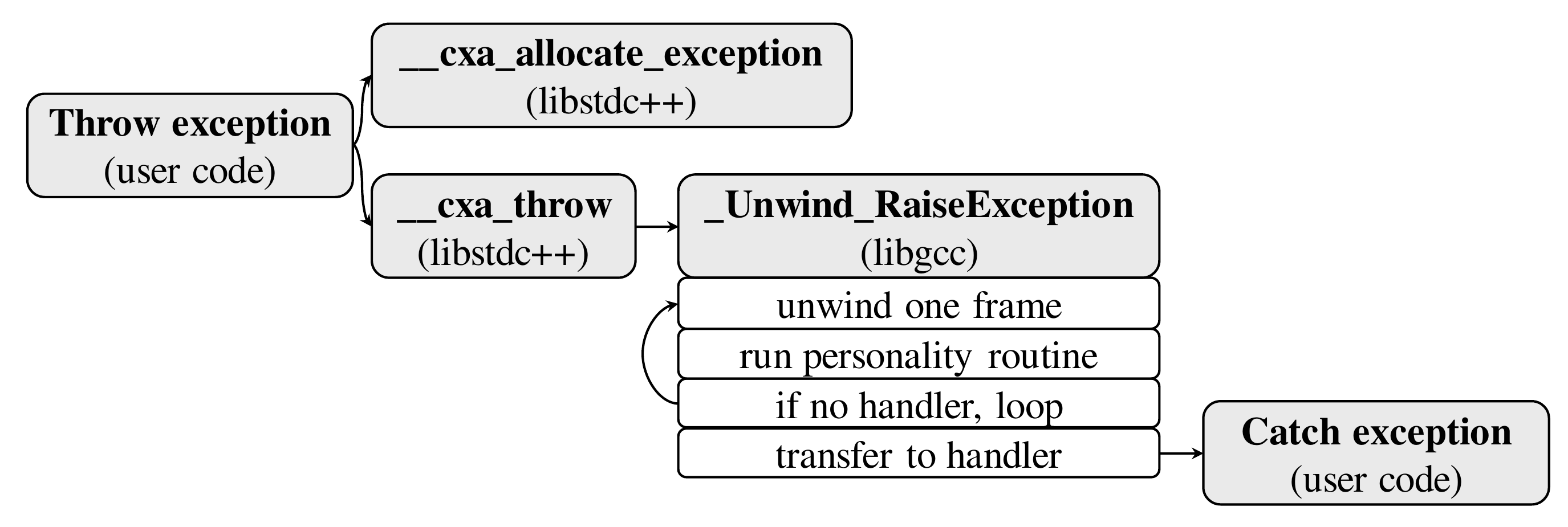}
  \caption{Workflow of exception handling in C++ programs.}
  \label{fig:eh-overview}
  \vspace{-0.8em}
\end{figure}

\section{Demystifying Exception Handling}
\label{sec:bcg-ehf}

In this section, we describe the technical details of exception handling at the binary level and unveil the types of exception handling information that can help function start detection. %We will start with the high level understanding and then delve into the details. 
% Our description will focus on x64 binaries that run on Linux platforms, to be aligned with our goals.

\subsection{Exception Handling at the High Level} Exception handling is the process of responding to the occurrence of exceptions during the execution of a program. Support of exception handling has become a standard feature of modern programming languages. For instance, C++ provides the \texttt{try}, \texttt{throw}, and \texttt{catch} clauses to facilitate handling of exceptions. To explain exception handling, we use the C++ example in Figure~\ref{fig:cppexc}. Exception handling in other programming languages follows a similar format, although using different grammar.

As shown in Figure~\ref{fig:cppexc}, the \texttt{main} function receives two integers from the user and attempts to divide them by calling \texttt{div}. In normal cases, \texttt{div} returns the division result to \texttt{main}, but if the divisor is zero, it throws an exception which will then be caught and handled by \texttt{main}. To realize exception handling in this case, execution has to go through two key steps. First, it needs to find the proper handler for the exception. As shown in our example, the throwing of an exception and the suitable handler for that exception can lie in different functions. As such, exception handling may need to search in the call chain on the stack, including the current function where the exception is thrown and all the caller functions. Second, after finding the proper handler, execution is redirected to it.
% the control flow switches from the exception occurrence to that handler and continues execution.

The above two steps are mainly completed by a special procedure called \emph{stack unwinding}. When an exception occurs, stack unwinding linearly searches every function on the call stack for the exception handler. While searching the exception handler, stack unwinding concurrently updates the stack by removing the stack frame of each searched function until the correct handler is identified. Following that, stack unwinding sets the stack pointer to the frame of the function with the correct handler, recovers the contexts in that function, and switches the execution to that handler.

In Figure~\ref{fig:cppexc}, once \texttt{div} throws the exception at line 3, the execution will in turn search \texttt{div} and \texttt{main} to locate the right handler at line 12 in \texttt{main}. In this process, the execution will remove \texttt{div}'s stack frame and then set the stack pointer to \texttt{main}'s frame. Finally, the execution will recover the context of \texttt{main} and switches to the \texttt{catch} clause at line 12.

\subsection{Exception Handling under the Hood}

In this section, we further reveal the under-the-hood mechanism of exception handling and stack unwinding.
We follow the same setting of our running example: exception handling in x64 binaries compiled from C++ programs.

Figure~\ref{fig:eh-overview} shows the workflow of the exception handling procedure. We will focus on the part of stack unwinding since other parts are not related to function detection. Stack unwinding is mainly completed by the \texttt{\_Unwind\_RaiseException} function from C library (libgcc). We describe how \texttt{\_Unwind\_RaiseException} performs the stack unwinding procedure as follows. For simplicity, we abbreviate \texttt{\_Unwind\_RaiseException} as $F_U$.

\begin{figure}[t!]
  \centering
  \includegraphics[width=0.95\linewidth]{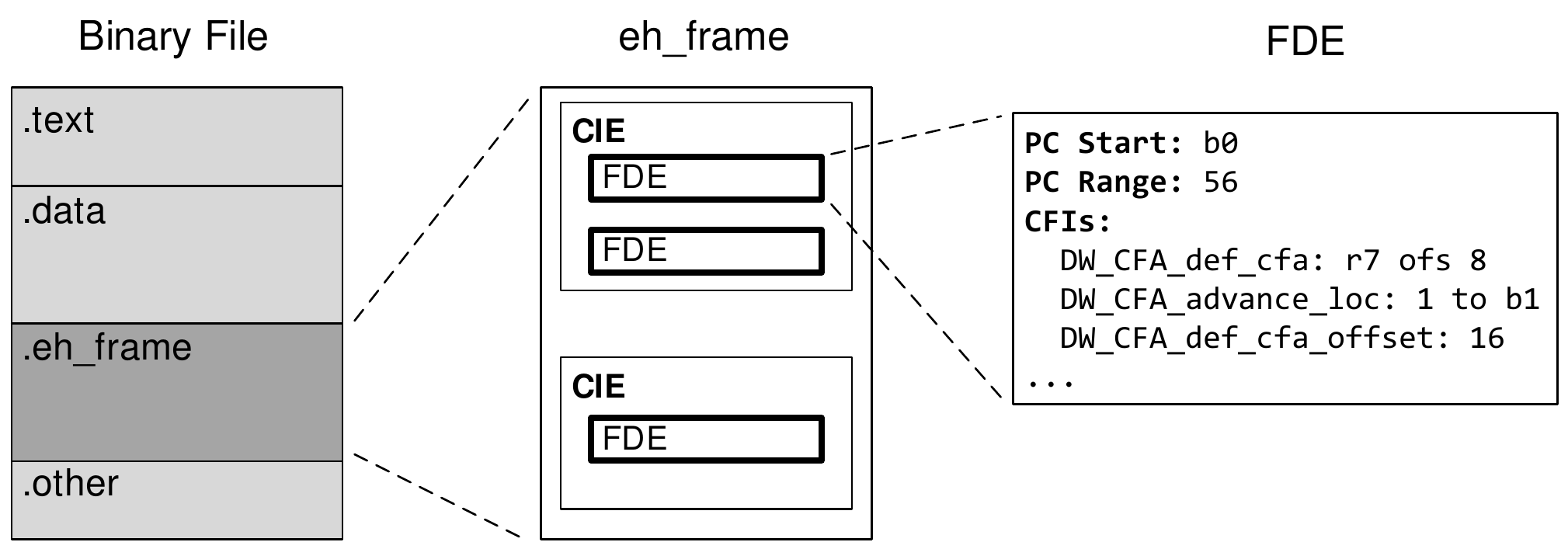}
  \caption{An overview of the \ehf section.}
  \label{fig:eh-frame}
  \vspace{-0.8em}
\end{figure}

\iffalse
\begin{table}[t!]
  \caption{Major fields in an FDE.}
  \label{tab:fde}
  \centering
  \scriptsize
  \begin{tabular}{l | c}
    \hline
    \hline
    \textbf{\small{Field}}           & \textbf{\small{Description}}                                               \\ \hline
    Length                   & Bytes of the CIE structure                                         \\ \hline
    CIE Pointer              & The start address of the CIE record                  \\ \hline
    Extended Length          & The length of the CIE structure                                    \\ \hline
    PC Begin                 & The begin location of the instructions in the FDE \\ \hline
    PC Range                 & The number of bytes in the FDE       \\ \hline
    Call Frame Instructions  & The records of changes to the call frame                        \\ \hline
    Augmentation Data        & A block of data defined by the CIE                                 \\ \hline
    Augmentation Data Length & The length of the augmentation data                                \\ \hline
    Padding                  & Extra bytes to align the FDE structure                             \\ \hline\hline
  \end{tabular}
\end{table}
\fi

\begin{figure*}[t!]
  \centering
  \begin{subfigure}[b]{0.28\textwidth}
    \centering
\begin{lstlisting}[language={[x86masm]Assembler},basicstyle={\scriptsize\ttfamily},numberstyle=\scriptsize,xleftmargin=5mm,deletekeywords={df,}]
; function start
b0: push   rbp        ; 1
b1: lea    rax,[rip+0x36d8b8]
b8: lea    rbp,[rdi+0x50]
bc: push   rbx        ; 2
bd: lea    rbx,[rdi+0xb0]
c4: sub    rsp,0x8    ; 3
c8: mov    QWORD PTR [rdi],rax
cb: nop    DWORD PTR [rax]
d0: sub    rbx,0x18
d4: mov    rdi, QWORD PTR [rbx]
d7: call   3d5c0 <qfree@plt>
dc: cmp    rbp,rbx
df: jne    45fd0 <main+0x55e0>
e1: add    rsp,0x8    ; 4
e5: pop    rbx        ; 5
e6: pop    rbp        ; 6
e7: ret
; function end
\end{lstlisting}
    \caption{Assembly code}
    \label{fig:sub1}
  \end{subfigure}
  \hspace{1mm}
  \begin{subfigure}[b]{0.35\textwidth}
    \centering
% cie = 001a8
\begin{lstlisting}[basicstyle={\scriptsize \ttfamily},numberstyle=\scriptsize,xleftmargin=5mm,xrightmargin=2mm,morekeywords={PC,Begin,Range,CFIs}]
00001070 FDE 
  PC Begin: b0
  PC Range: 56
  CFIs:
    DW_CFA_def_cfa: r7 (rsp) ofs 8
    DW_CFA_advance_loc: 1 to b1  // 1
    DW_CFA_def_cfa_offset: 16
    DW_CFA_offset: r6 (rbp) at cfa-16
    DW_CFA_advance_loc: 12 to bd  // 2
    DW_CFA_def_cfa_offset: 24
    DW_CFA_offset: r3 (rbx) at cfa-24
    DW_CFA_advance_loc: 11 to c8  // 3
    DW_CFA_def_cfa_offset: 32
    DW_CFA_advance_loc: 29 to e5  // 4
    DW_CFA_def_cfa_offset: 24
    DW_CFA_advance_loc: 1 to e6  // 5
    DW_CFA_def_cfa_offset: 16
    DW_CFA_advance_loc: 1 to e7  // 6
    DW_CFA_def_cfa_offset: 8
\end{lstlisting}
    \caption{FDE entry from \ehf}
    \label{fig:sub2}
  \end{subfigure}
  \hspace{1mm}
  \begin{subfigure}[b]{0.33\textwidth}
    \centering
    \includegraphics[width=\textwidth]{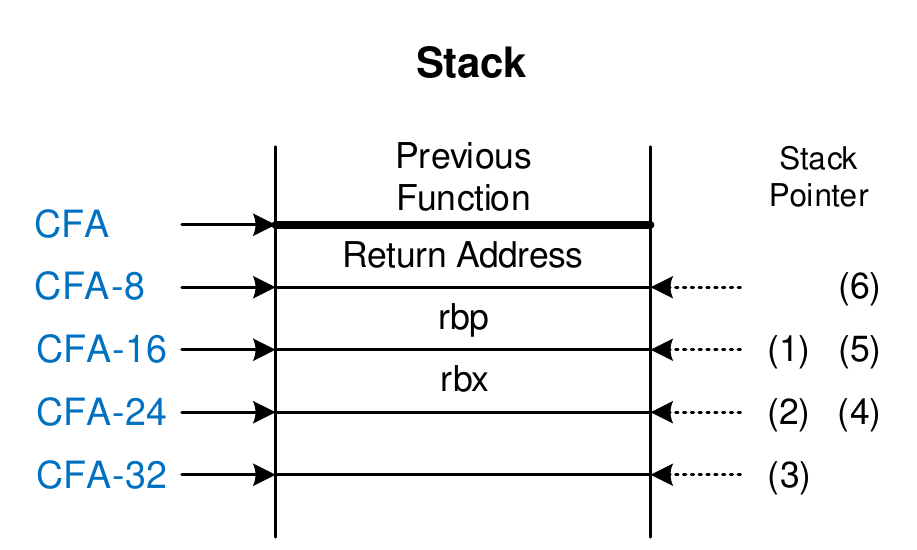}
    \vspace{1.5cm}
    \caption{The run-time stack}
    \label{fig:sub3}
  \end{subfigure}
  \caption{A function from IDA-Pro 7.2 and its FDE. Addresses of instructions are simplified to only keep the lower two digits.}
  \label{fig:eh-example}
  \vspace{-0.8em}
\end{figure*}

\begin{description}
\item[\ding{182}] $F_U$ first checks the program counter (PC), \ie the \texttt{rip} register, at the {\tt throw} statement and determines the current function (\eg \texttt{div} in Figure~\ref{fig:cppexc}) based on the PC.

\item[\ding{183}] $F_U$ then checks if the current function has a proper handler. Specifically, it checks whether the current function has a \texttt{catch} block that can handle the \texttt{throw}. If a proper \texttt{catch} block is found, $F_U$ switches the PC to the \texttt{catch} block. Otherwise, $F_U$ recovers the registers saved by the current function and destroys its stack frame by adjusting the stack pointer (SP). Then, $F_U$ goes to the next step.

\item[\ding{184}] $F_U$ finds the caller function on the stack (\eg \texttt{main} in Figure~\ref{fig:cppexc}) and repeats \ding{183}, using the return address as the new PC. However, if the stack frame is empty, $F_U$ will invoke \texttt{terminate} to make the program exit abnormally.

\end{description}

As unveiled by the description above, \ding{182}-\ding{184} critically depend on three tasks: ($\mathbb{T}_1$) given PC, finding the function containing the PC; ($\mathbb{T}_2$) given PC and the corresponding SP, determining the call frame of the current function and its return address; ($\mathbb{T}_3$) given PC and the corresponding SP, recovering the registers saved by the current function. To complete the three tasks, $F_U$ leverages information from a special section called \ehf, which is also the key data empowering function detection. In the rest of this section, we will give a brief introduction of \ehf and then explain how it helps complete the three tasks.

\subsection{EH\_Frame: Key Data Structure for Exception Handling}
\label{subsec:ehframe}

\myparatight{Overview of EH\_Frame} As illustrated in Figure~\ref{fig:eh-frame}, \ehf is a separate section in a binary file. It is structured as a list of Common Information Entries (CIE), each corresponding to an object file linked into the binary file. A CIE carries one or more Frame Description Entries (FDE), and typically, one FDE records the information of a unique function from the CIE's object file.

\myparatight{Exception Handling with EH\_Frame} The major information in \ehf used by exception handling resides in the FDEs. An FDE record consists of a list of fields, among which \texttt{PC Begin}, \texttt{PC Range}, and \texttt{Call Frame Instructions} (CFIs) are indispensable to tasks $\mathbb{T}_1$-$\mathbb{T}_3$. In the following, we will follow the example in Figure~\ref{fig:eh-example} to explain how the three fields are used to complete $\mathbb{T}_1$-$\mathbb{T}_3$.

Figure~\ref{fig:sub1} shows the assembly code of a function extracted from IDA-Pro 7.2 and Figure~\ref{fig:sub2} shows the corresponding FDE. Line 2 and 3 in Figure~\ref{fig:sub2} presents the {\tt PC Begin} and the {\tt PC Range} fields in the FDE. They explicitly give the start address and length of the function body. \emph{Using the PC information in the FDEs, exception handling can easily find the function containing a given PC, thus completing task $\mathbb{T}_1$}. 

% Introduce CFA
The rest part of Figure~\ref{fig:sub2} (line 4-19) presents \texttt{CFIs}, a group of special instructions describing the unwinding rules. Due to historical reasons, the format of \texttt{CFI} follows the DWARF standard~\cite{dwarf2010dwarf}. The core concept introduced by these unwinding rules is called ``Canonical Frame Address (CFA)''. CFA is a universal variable that refers to the base address of the current stack frame (typically the highest address), which helps to uniform the various representations of the frame pointer introduced by compilers.
% The benefit of using CFA is to uniform the various representations of the frame pointer introduced by compilers.
%In Figure~\ref{fig:sub2}, CFA is represented by the stack pointer \texttt{rsp}.
There are four main types of instructions involved in the unwinding rules:
\begin{itemize}
\item \texttt{DW\_CFA\_def\_cfa} defines how CFA is represented, normally in the format of an offset to a designated register.
\item \texttt{DW\_CFA\_advance\_loc} records the location of an instruction that changes the register used to represent the CFA or saves certain registers to the stack. The instruction location is represented by an offset to the function start.
\item \texttt{DW\_CFA\_def\_cfa\_offset} describes the rule to calculate CFA when the value of the register representing CFA is changed. The rule typically follows the format of an offset relative to that register.
\item \texttt{DW\_CFA\_offset} records the saving of certain registers to the stack, covering both the number and the location of the saved register.
\end{itemize}

We continue using the example in Figure~\ref{fig:eh-example} to explain how the above instructions describe concrete unwinding rules. At line 5 in Figure~\ref{fig:sub2}, a \texttt{DW\_CFA\_def\_cfa} instruction defines that \texttt{rsp} is used to represent the CFA and initially, \texttt{CFA = rsp + 8}. Across the entire function, there are six instructions changing \texttt{rsp}, respectively marked as 1-6 in the comments in Figure~\ref{fig:sub1}. Correspondingly, FDE records each change with a separate \texttt{DW\_CFA\_advance\_loc} instruction, also marked as 1-6 in the comments in Figure~\ref{fig:sub2}. Following each \texttt{DW\_CFA\_advance\_loc} instruction, FDE appends a \texttt{DW\_CFA\_def\_cfa\_offset} instruction to describe how to re-calculate the CFA. Consider line 2 in Figure~\ref{fig:sub1} as an example. The instruction pushes \texttt{rbp} to the stack, decreasing \texttt{rsp} by 8. Accordingly, line 6 in Figure~\ref{fig:sub2} indicates that the instruction before address \texttt{b1} makes a change to the register representing the CFA (\ie \texttt{rsp}); line 7 in Figure~\ref{fig:sub2} describes that now the offset between CFA and \texttt{rsp} is 16 bytes, namely \texttt{CFA = rsp + 16}. The rest five \texttt{DW\_CFA\_def\_cfa\_offset} instructions can be interpreted similarly and the run-time stack is shown in Figure~\ref{fig:sub3}.

\vspace{-0.05em}
The above mechanism guarantees that, given PC and {\tt rsp} at any execution point, CFA can be correctly calculated. This essentially ensures that (i) the range of the current stack frame can be determined since CFA always refers to the highest address of the current stack frame and (ii) the return address of the current function can be found because return address is located right below the top of the current stack frame (\ie at \texttt{CFA - 8}). \emph{Therefore, using information in the CFI, exception handling can correctly complete task $\mathbb{T}_2$}. 

Referring back to the example in Figure~\ref{fig:eh-example}, the first instruction in Figure~\ref{fig:sub1} pushes \texttt{rbp}, a callee-saved register, to the stack. Correspondingly, FDE inserts a \texttt{DW\_CFA\_offset} instruction at line 8 in Figure~\ref{fig:sub2}, indicating that the instruction before address \texttt{b1} saves register number 6 (\ie \texttt{rbp}) to address \texttt{CFA - 16} on the stack. Similar to this case, whenever a callee-saved register is stored on the stack, FDE inserts a \texttt{DW\_CFA\_offset} instruction. As such, \emph{given PC and \texttt{rsp} at any point of the execution, exception handling can learn what callee-saved registers exist in the current stack frame, and thus, complete $\mathbb{T}_3$ in the unwinding process}.

We also note that x64 binaries compiled from C programs similarly carry FDEs, as verified by our studies in \S~\ref{sec:study-cov}. 
In fact, the x64 ABI mandates FDEs in such binaries as many library functions like {\tt backtrace, \_\_builtin\_return\_address} need FDEs to support their functionality. 
% On the one hand, the x64 ABI mandates FDEs in such binaries. On the other hand, many library functions like {\tt backtrace} need FDEs to support their functionality.

%\vspace{1ex}
%\begin{mdframed}
%\noindent\textbf{Summary:} FDEs are indispensable structures for exception %handling, which give critical information about functions, such as %function starts and function ranges.
%\end{mdframed}

% \section{Optimizing Coverage of Function Start Detection with Call Frames}
\section{Exploring Coverage of Function Start Detection with Call Frames}
\label{sec:study-cov}

In this section, we aim at our first research goal --- \emph{exploring the best strategies that use call frames towards optimal coverage of function starts with a minimal harm to the reliability}. We will study the detection of function starts using FDEs with both safe and unsafe approaches from existing tools, and hence, understand which combination of approaches can bring the best balance between coverage and risks. To be more specific, we will center around three questions: 

\begin{description}
\setlength\itemsep{0em}
    \item [$\mathbb{Q}_1$ -] {\em Using only FDEs, how many function starts can be detected?} 
    \item [$\mathbb{Q}_2$ -] {\em Using FDEs and safe approaches, particularly recursive disassembly, how many function starts can be detected?} 
    \item [$\mathbb{Q}_3$ -] {\em Can unsafe approaches, such as the heuristics used by existing tools, help detect more function starts? What are their side effects?}
\end{description}

To answer the above questions, we perform a set of empirical studies with a large corpus of x64 binaries as follows.

\begin{table}[t!]
% \begin{minipage}{1\linewidth}
\setlength\tabcolsep{4pt}
  \centering
  \caption{Wild binaries in our study. \textbf{\small Open} - open source or not; \textbf{\small EHF} - having \ehf or not; \textbf{\small Sym} - having symbols or not;
  \textbf{\small FDE} - ratio of functions with FDEs (v.s. symbols).} 
  \label{tab:comtools}
  \scriptsize
  \begin{tabular}{l c c c c c}
    \hline\hline
    \textbf{\small Software}       & \textbf{\small Open} & \textbf{\small EHF} & \textbf{\small Sym} & \textbf{\small FDE} & \textbf{\small Note} \\ \hline
    Atom-1.49.0           &\cmark      & \cmark    & \xmark & \textthreequartersemdash & gcc-7.3.0;c++ \\ 
    Simplenot-1.4.13      & \cmark     & \cmark    & \xmark  & \textthreequartersemdash &gcc-4.6.3;c++ \\ 
    % Simplenot-1.4.13      & \cmark     & \cmark    & \cmark & 95.30 & 97.49 &gcc-4.6.3 \\
    % OpenShot-2.4.4        & \cmark     & \cmark    & \cmark & 97.92 & 100.0 & gcc-4.8.4  \\ 
     OpenShot-2.4.4        & \cmark     & \cmark    & \xmark & \textthreequartersemdash & gcc-4.8.4; c  \\ 
    %uget-gtk-2.2.1        & \cmark     & \cmark    & \cmark   \\ 
    % seamonkey-2.49.5      & \cmark     & \cmark    & \cmark  & 97.77 & 98.22 & gcc-4.8.5 \\ 
    seamonkey-2.49.5      & \cmark     & \cmark    & \xmark  & \textthreequartersemdash & gcc-4.8.5; c++ \\ 
    
    % lmms-1.2.1            & \cmark     & \cmark    & \cmark & gcc-5.3.0  \\ 
    % mupdf-1.16.1          & \cmark     & \cmark    & \cmark  & 98.69 & 98.74 & gcc-7.4.0 \\ 
    mupdf-1.16.1          & \cmark     & \cmark    & \xmark   & \textthreequartersemdash & gcc-7.4.0; c \\
    % chromium-83.0.4103.61   & \xmark     & \cmark    & \cmark   &  & & gcc-7.5.0 \\
    % launchy-2.5.1         & \cmark     & \cmark    & \cmark  & 54.09 & 53.09 & \textthreequartersemdash \\ 
    % laverna-0.7.1         & \cmark     & \cmark    & \cmark & 96.93 & 97.67 & gcc-4.6.3  \\ 
    laverna-0.7.1         & \cmark     & \cmark    & \xmark  & \textthreequartersemdash & gcc-4.6.3; c++  \\ 
    franz-5.4.0           & \cmark     & \cmark    & \xmark  & \textthreequartersemdash & gcc-4.6.3; c++ \\ 
    
    % Nightingale-1.12.1    & \cmark     & \cmark    & \cmark  & 95.30 & 95.63 & gcc-4.7.2 \\ 
    Nightingale-1.12.1    & \cmark     & \cmark    & \xmark   & \textthreequartersemdash & gcc-4.7.2; c \\ 
    palemoon-28.8.0       & \cmark     & \cmark    & \xmark   & \textthreequartersemdash  & c++ \\
    % palemoon-28.8.0       & \cmark     & \cmark    & \cmark   & 96.71 & 96.95 & \textthreequartersemdash \\
   
    evince-3.34.3         & \cmark     & \cmark    & \xmark  & \textthreequartersemdash  & c \\ 
    %  evince-3.34.3         & \cmark     & \cmark    & \cmark  & 97.60 & 97.71 & \textthreequartersemdash \\ 
    
    % amarok-2.9.0            & \cmark     & \cmark    & \cmark & 90.01 & 100.0 & \textthreequartersemdash  \\
    amarok-2.9.0            & \cmark     & \cmark    & \xmark & \textthreequartersemdash  & c  \\ 
   
    % Brave-Browser-1.1.20  & \cmark     & \cmark    & \cmark   \\ 
    % deadbeef-1.8.2        & \cmark     & \cmark    & \cmark & 95.98 & 99.16 & \textthreequartersemdash  \\
    deadbeef-1.8.2        & \cmark     & \cmark    & \xmark & \textthreequartersemdash  & c  \\
    % qBittorrent-4.2.5        & \cmark     & \cmark    & \cmark & 85.24 & 86.83 & \textthreequartersemdash  \\
    qBittorrent-4.2.5        & \cmark     & \cmark    & \xmark & \textthreequartersemdash   & c++  \\
    % pdftex-3.14159265        & \cmark     & \cmark    & \cmark & 98.04 & 98.24 & \textthreequartersemdash  \\
    pdftex-3.14159265 & \cmark     & \cmark    & \xmark & \textthreequartersemdash  & c  \\
    eclipse-4.11 & \cmark     & \cmark    & \xmark & \textthreequartersemdash & gcc-4.8.5; c  \\
     VS Code-1.40.2        & \cmark     & \cmark    & \xmark  & \textthreequartersemdash  & gcc-7.3.0; c++ \\
     VirtualBox-5.2.34     & \cmark     & \cmark    & \cmark   & 100.0  & c++ \\
    gv-3.7.4              & \cmark     & \cmark    & \cmark  & 100.0  & c \\ 
    okular-1.3.3            & \cmark     & \cmark    & \cmark   & 100.0 & c++ \\ 
    gcc-7.5        & \cmark     & \cmark    & \cmark & 100.0  & c  \\
    
    wkhtmltopdf-0.12.4     & \cmark     & \cmark    & \cmark   & 100.0  & c \\ 
    firefox-78.0.2  & \cmark     & \cmark    & \cmark & 100.0  & c++ \\ 
    qemu-system-2.11.1    & \cmark     & \cmark    & \cmark  & 100.0  & c \\ 
    
    ThunderBird-68.10.0    & \cmark     & \cmark    & \cmark  & 100.0  &  gcc-6.4.0; c++ \\
   
    Smuxi-Server          & \cmark     & \cmark    & \cmark  & 100.0  & gcc-5.3.1; c  \\
    % eclipse-4.11          & \cmark     & \cmark    & \xmark   \\ 
    \hline
    % TeamViewer-15.0.8397  & \xmark     & \cmark    & \cmark   & 71.23 & 72.02 & gcc-7.2.0 \\ 
    TeamViewer-15.0.8397  & \xmark     & \cmark    & \xmark   & \textthreequartersemdash  & gcc-7.2.0; c++ \\
    skype-8.55.0.141      & \xmark     & \cmark    & \xmark & \textthreequartersemdash  & gcc-7.3.0; c++   \\ 
    % trillian-6.1.0.5      & \xmark     & \cmark    &  \cmark &  63.82 & 64.93 & \textthreequartersemdash  \\ 
    trillian-6.1.0.5      & \xmark     & \cmark    &  \xmark &  \textthreequartersemdash  & c++  \\ 
    opera-65.0.3467.69    & \xmark     & \cmark    & \xmark & \textthreequartersemdash  & gcc-7.3.0; c++   \\ 
    yandex-browser-19.12.3  & \xmark     & \cmark    & \xmark &   \textthreequartersemdash  & gcc-7.3.0; c++ \\ 
    % SpiderOakONE-7.5.01   & \xmark     & \cmark    & \cmark & 99.15 & 99.15  & gcc-4.1.2  \\
    SpiderOakONE-7.5.01   & \xmark     & \cmark    & \xmark & \textthreequartersemdash  & gcc-4.1.2; c  \\
    % vivaldi-2.10.1745     & \xmark     & \cmark    & \cmark &  & & gcc-7.5.0  \\
    slack-4.2.0           & \xmark     & \cmark    & \xmark & \textthreequartersemdash  & gcc-7.3.0; c++   \\
    rainlendar2-2.15.2    & \xmark     & \cmark    & \xmark & \textthreequartersemdash  & gcc-5.4.0; c++   \\
    %googleearth-7.3.2     & \xmark     & \cmark    & \cmark & gcc-4.9.4  \\
    % chrome-79.0.3945.88   & \xmark     & \cmark    & \cmark   &  & & gcc-7.5.0 \\
    sublime-3211          & \xmark     & \cmark    & \xmark   & \textthreequartersemdash  & gcc-6.3.0; c++ \\
    % Discord-9.0.9         & \xmark     & \cmark    & \cmark  & gcc-7.3.0 \\
    netease-cloud-music-1.2.1      & \xmark     & \cmark    & \xmark  & \textthreequartersemdash  &  c++ \\
    wps-11.1.0.8865       & \xmark     & \cmark    & \xmark  & \textthreequartersemdash  & c++ \\
    wpp-11.1.0.8865       & \xmark     & \cmark    & \xmark   & \textthreequartersemdash  & c++ \\
    wpspdf-11.1.0.8865    & \xmark     & \cmark    & \xmark  & \textthreequartersemdash  & c++ \\
    wpsoffice-11.1.0.8865 & \xmark     & \cmark    & \xmark  & \textthreequartersemdash  & c++  \\
    ida64-7.2             & \xmark     & \cmark    & \xmark   & \textthreequartersemdash  & gcc-4.8.2; c++ \\
    zoom-7.19.2020  & \xmark     & \cmark    & \xmark  & \textthreequartersemdash  &  gcc-4.8.5; c++ \\
    binaryninja-1.2       & \xmark     & \cmark    & \cmark  & 100.0  & gcc-5.4.0; c++\\ 
    FoxitReader-4.4.0911  & \xmark     & \cmark    & \cmark  & 99.99  & gcc-4.8.4; c++ \\\hline 
    \textbf{Avg.} & - & -  & -  & \textbf{99.99} & - \\
   
    \hline\hline
  \end{tabular}
%   \end{minipage}
\vspace{-1.5em}
\end{table}

\subsection{Setup of Studies} 
\label{subsec:study-setup}

\subsubsection{Preparation of Datasets} 

We collected two sets of x64 binaries, one from the wild and one built from source code.

%18 close-source binaries and 25 pre-built binaries from open-source programs, all in their latest versions. Details of the binaries are shown in Table~\ref{tab:comtools}. These binaries cover nearly all the common software we use in our daily life, ranging from editors to browsers and tele-conference clients. 

\myparatight{Dataset 1} The first dataset are binaries collected from the wild, including 18 close-source binaries and 25 pre-built binaries from open-source programs. Details of the binaries are shown in Table~\ref{tab:comtools}. The binaries cover nearly all the common types of software we use in our daily life, ranging from editors to browsers and tele-conference clients. The binaries also cover both C programs and C++ programs.

\begin{table}[t]
  \centering
  \caption{Self-built programs used in our study. \textbf{\small EHF} - having \ehf or not; \textbf{\small FDE} - the ratio of functions that have FDEs, where the baseline is symbols.}
  \label{tab:dataset}
  \scriptsize
  \begin{tabular}{m{0.09\textwidth} m{0.04\textwidth} c c c c}
    \hline\hline
    \textbf{\small Project} & \textbf{\small Type} & \textbf{\small \# Prog/Bins} & \textbf{\small EHF} & \textbf{\small FDE} & \textbf{\small Lang} \\ \hline
    % Unzip-6.0           & Utilities     &               &                      & \cmark    & \cmark   \\ 
    Coreutils-8.30      & Utilities     &    105/840           &    \cmark    & 100.0  & C \\ 
    % 7-zip-19            & Utilities     &               &                      & \cmark    & \cmark   \\ 
    Findutils-4.4       & Utilities     &    3/24           & \cmark                 & 100.0  & C  \\ 
    Binutils-2.26       & Utilities     &   17/136          &      \cmark      & 100.0 &  C/C++ \\ \hline
    % Tiff-4.0            & Utilities     &               &                      & \cmark    & \cmark   \\ 
    Openssl-1.1.0l      & Client        &     1/4          &    \cmark      & 96.40 & C  \\ 
    % Putty-0.73          & Client        &               &                      & \cmark    & \cmark   \\ 
    D8-6.4              & Client        &     1/4            &  \cmark     & 100.0 & C++ \\ 
    % Filezilla-3.44.2    & Client        &               &                      & \cmark    & \cmark   \\ 
    Busybox-1.31        & Client        &     1/8  &   \cmark  & 100.0   & C \\ 
    Protobuf-c-1        & Client        &     1/6          &  \cmark    &  100.0  & C++ \\ 
    ZSH-5.7.1           & Client        &     1/2            &  \cmark   & 100.0  & C  \\ 
    % VIM-8.1             & Client        &               &                      & \cmark    & \cmark   \\ 
    % XML2-2.9.8          & Client        &               &  \cmark                    & \cmark    &    \\ 
    Openssh-8.0         & Client        &      7/28         &   \cmark   & 100.0 & C   \\ 
    Mysql-5.7.27            & Client        &       1/6        &  \cmark    & 100.0 & C++   \\
    Git-2.23            & Client        &       1/8        &  \cmark   & 100.0 & C \\
    filezilla-3.44.2            & Client        &       1/4        &  \cmark   & 100.0 & C++
    \\ \hline
    Lighttpd-1.4.54     & Server        &       1/8        &  \cmark    &  100.0 & C  \\ 
    Mysqld-5.7.27       & Server        &       1/6        &  \cmark    & 100.0 & C++   \\ 
    Nginx-1.15.0        & Server        &       1/6        &  \cmark   &  98.97 & C \\\hline
    % SQLite-3.32.0       & Server        &               &  \cmark                   & \cmark    &    \\
    Glibc-2.27          & Library       &       1/3            & \cmark     &  99.97 & C \\ 
    libpcap-1.9.0       & Library       &       1/8        &  \cmark    & 100.0 & C   \\ 
    libv8-6.4           & Library       &       1/4        &  \cmark   &  100.0 & C++ \\ 
    libtiff-4.0.10      & Library       &       1/8        & \cmark    & 100.0 & C   \\ 
    libxml2-2.9.8       & Library       &       1/8        & \cmark   &  100.0 & C  \\ 
    % libsqlite-3.32.0    & Library       &               & \cmark                     & \cmark    &    \\ 
    libprotobuf-c-1 & Library       &      1/8         & \cmark    &  100.0  & C++\\ \hline
    SPEC CPU2006        & Benchmark     &    30/223           & \cmark    &  99.99  & C/C++\\ \hline
    \textbf{Total}               &      -         &   \textbf{179/1352}            &        -              &      \textbf{99.87}     &  \textbf{-}        \\ \hline\hline
  \end{tabular}
  \vspace{-1.5em}
\end{table}

\myparatight{Dataset 2} The second dataset is compiled from widely-used open source programs. As shown in Table~\ref{tab:dataset}, the dataset includes 179 programs used by a recent study~\cite{sok:sp20}, covering both applications and libraries that are written in C/C++. These programs carry highly diverse functionality and complexity; They also contain both hand-written assembly code and hard-coded machine code. To further increase the diversity, we compiled the 179 programs into x64 binaries with both LLVM (version 6.0.0) and GCC (version 8.1.0), using optimization level O2, O3, Ofast, and Os. We omitted O0 and O1 since they are not widely used in practice. At the end, we produced 1,352 binaries in total. 
% , less than the theoretically possible 1,432 because some programs cannot compile with certain configurations.

%\begin{figure}[t]
%    \centering
%    \includegraphics[scale=0.436]{figure/ccr.pdf}
%    \caption{The frameworks to obtain ground truth of functions. The boxes in gray color indicate components that are modified in LLVM, GCC, GAS, and
%    Gold Linker.}
%    \label{fig:gt}
%\end{figure}

\subsubsection{Generation of Ground Truth} 
To measure the detection results, ground truth about the function starts is required. One common approach to obtaining the ground truth is to use the symbols, but we found that symbols are not perfect: (i) the symbols for hundreds of destructor functions in our dataset are missing; (ii) the symbols for a small group of assembly functions in our dataset have incomplete types. More importantly, symbols can introduce a significant group of false positives, as we will show in \S~\ref{sec:study-accuracy}. Thus, we only use symbols for the pre-compiled binaries in dataset 1 since we have no other options; while for dataset 2, we use a compiler-based approach to produce more complete and more accurate ground truth. Specifics are as follows.

\myparatight{Ground Truth for Dataset 1} As described above, we considered symbols as the ground truth of function starts for the binaries from the wild. Specifically, among the 25 binaries pre-built from open source projects, we found symbols in 1 of them and we successfully installed symbols for 8 others. For the closed-source binaries, we found symbols in 2 of them. In total, we obtained symbol-based ground truth for 11 wild binaries and our studies only considered them.

%In this step, we only consider binaries that have symbols because we use symbols as the baseline. Among the 25 open-source binaries, we found one is not stripped and we also manually installed symbols for 8 other ones. For the closed-source binaries, we only included 2 that are not stripped.

\myparatight{Ground Truth for Dataset 2}
To generate a better ground truth than the symbols, we re-used the frameworks developed by~\cite{sok:sp20} to intercept the end-to-end compiling and linking process to obtain all function starts. 
% Technically, the frameworks follow the idea of CCR~\cite{koo2018compiler} to intercept the LLVM compiler. CCR extends the LLVM Machine Code (MC) layer. When assembling a bitcode/assembly file to an object file, the extended MC collects functions. To concatenate information from multiple object files, CCR further instruments the  GNU Gold Linker to merge those items during linking. The frameworks also port the idea of CCR to GCC by instrumenting the RTL pass in GCC to label functions in the assemble code. The frameworks further customize GNU Assembler (GAS) to record the locations of functions in the object files and reuse the CCR linker to merge object files. 

\subsection{Answering Question $\mathbb{Q}_1$}
\label{subsec:recursive}

\myparatight{Comparing with Symbols} In our first study, we extracted the {\tt PC Begin} fields from all FDEs and compared them with symbols. Not surprisingly, FDEs and symbols are highly overlapped. In the 11 wild binaries, FDEs cover 101,882 (99.99\%) of the 101,891 symbols. In 9 out of the 11 binaries, FDEs cover all the symbols (see the column of {\bf FDE} in Table~\ref{tab:comtools}). The results with self-built binaries are similar. In the 1,352 self-built binaries, FDEs cover 1,138,601 (99.87\%) of the 1,140,047 symbols. In 1,319 out of these binaries, FDEs cover all the symbols (see the column of {\bf FDE} in Table~\ref{tab:dataset}).

\myparatight{Comparing with Ground Truth} We further considered the {\tt PC Begin} fields in our self-built binaries as function starts and compared them with the compiler-generated ground truth. In total, the FDEs cover 1,103,832 of the 1,105,278 function starts. Despite the high overall coverage rate (99.87\%), {\em FDEs alone can still leave large coverage gaps in many 
binaries}. To be specific, FDEs miss function starts in 33 of our self-built binaries and the average number of missed functions is 43.82. 
In the binary built from Openssl with Ofast, FDEs miss 237 functions.
% , amounting to 3.69\% of the total functions. 

Through manual analysis, we found that the majority (1,330 out of 1,446) of the functions missed by FDEs are assembly functions. In principle, to comply with the ABI and generate FDE for every assembly function, the developers should write CFI directives~\cite{gas:online} manually. However, this only happens in infrastructural projects\footnote{Example: \url{https://github.com/openssl/openssl/blob/33388b44b67145af2181b1/crypto/aes/asm/aes-x86\_64.pl\#L608}} instead of everywhere, due to the complexity and error-proneness of manually creating unwinding rules. The other functions missed by FDEs are the instances of {\tt \_\_clang\_call\_terminate}, which are statically linked into the binaries by the Clang compiler.

% We also found that FDEs are not perfectly accurate. In 488 out of the 1,352 self-built binaries, there are FDEs whose {\tt PC Begin} does not point to a function start. When only considering the 488 binaries, the average number of wrong function starts introduced by FDEs is 71.25. In the binary compiled from Mysqld with Ofast, FDEs introduced 3,616 false function starts. We will separately discuss the errors in \S~\ref{sec:study-accuracy}.

%mysqld-gcc_Of

%# of binaries: 1352
%# of full coverage: 1319
%# of full accuracy: 864

%handwritten assembly code: 1330
%__clang_call_terminate: 116

%In principle, to comply with the ABI, the developers should manually insert an FDE with unwinding rules into assembly functions. However, this is only happening in infrastructural projects (\eg \url{https://github.com/openssl/openssl/blob/master/crypto/aes/asm/aesni-x86_64.pl#L619}).

%\vspace{1ex}
%\begin{mdframed}
%\noindent\textbf{Summary:} Overall, FDEs provide very high coverage of %function starts. However, only using FDEs can still result in large %coverage gaps, particularly when the program contains many assembly %functions.
%\end{mdframed}

\subsection{Answering Question $\mathbb{Q}_2$}
\label{subsec:recdias}

Following our first study, we then investigate whether recursive disassembly, a widely-used safe approach, can detect the missing function starts. Specifically, we ran the built-in recursive disassembly in both \angr and \ghidra, starting from addresses carried by FDEs and symbols. In the course of recursive disassembly, both \angr and \ghidra consider targets of call instructions as new function starts. As we focus on the effectiveness of recursive disassembly in this study, we disabled the extra heuristics used by \angr and \ghidra for function detection, including the tail call detection used by both tools, the function matching used by both tools, and the linear scan used by \angr (more details can be found in~\cite{sok:sp20}). To guarantee the accuracy of the experiment result, we only considered the self-built binaries since we have the precise ground truth for them.

\begin{figure*}[t!]
  \centering
  \begin{subfigure}[t]{0.32\textwidth}
    \centering
    \includegraphics[scale=0.26]{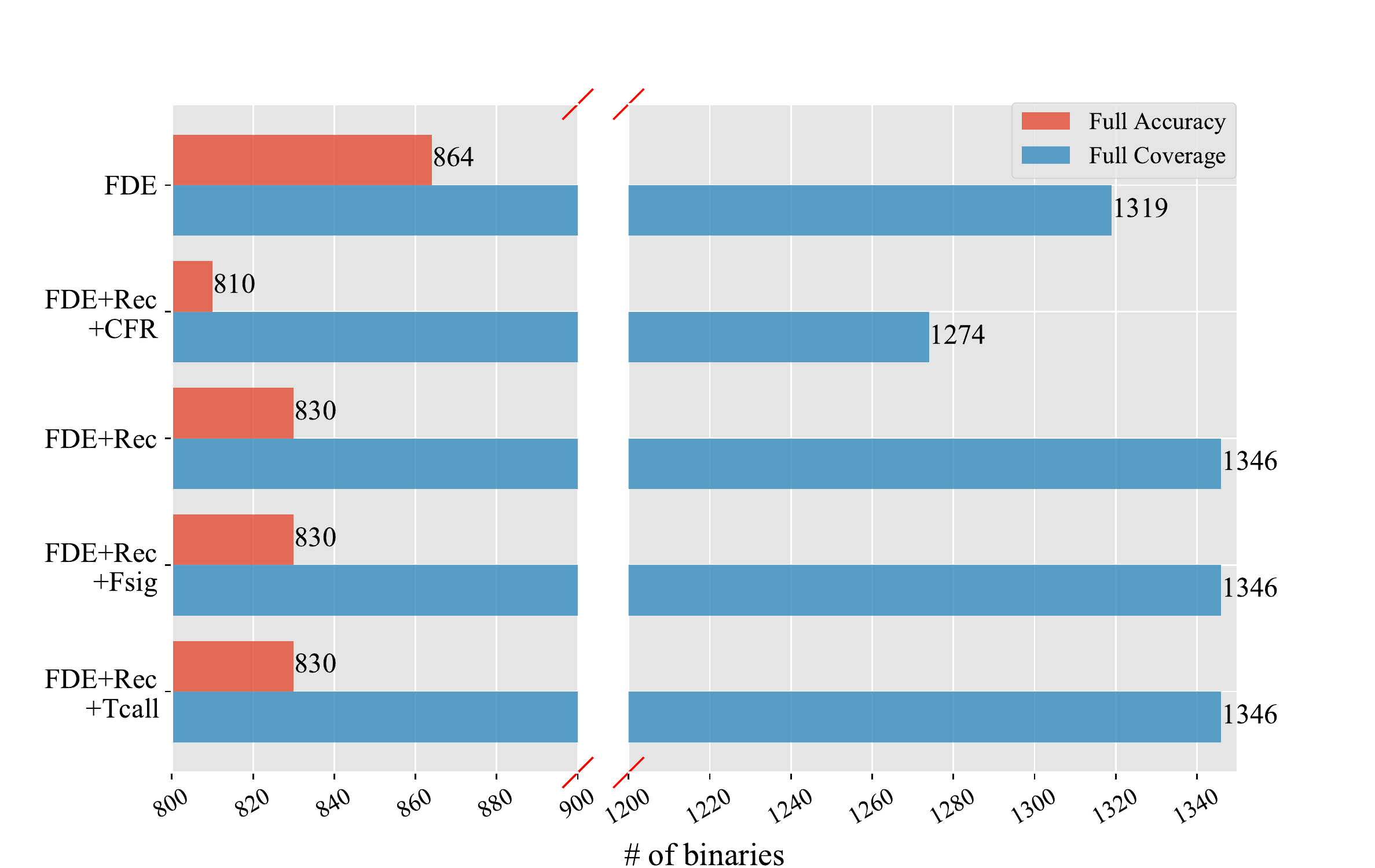}
    \caption{\ghidra (1,352 binaries)}
    \label{fig:full_cov_acc_ghidra}
    \end{subfigure}
  \begin{subfigure}[t]{0.32\textwidth}
  \centering
  \includegraphics[scale=0.26]{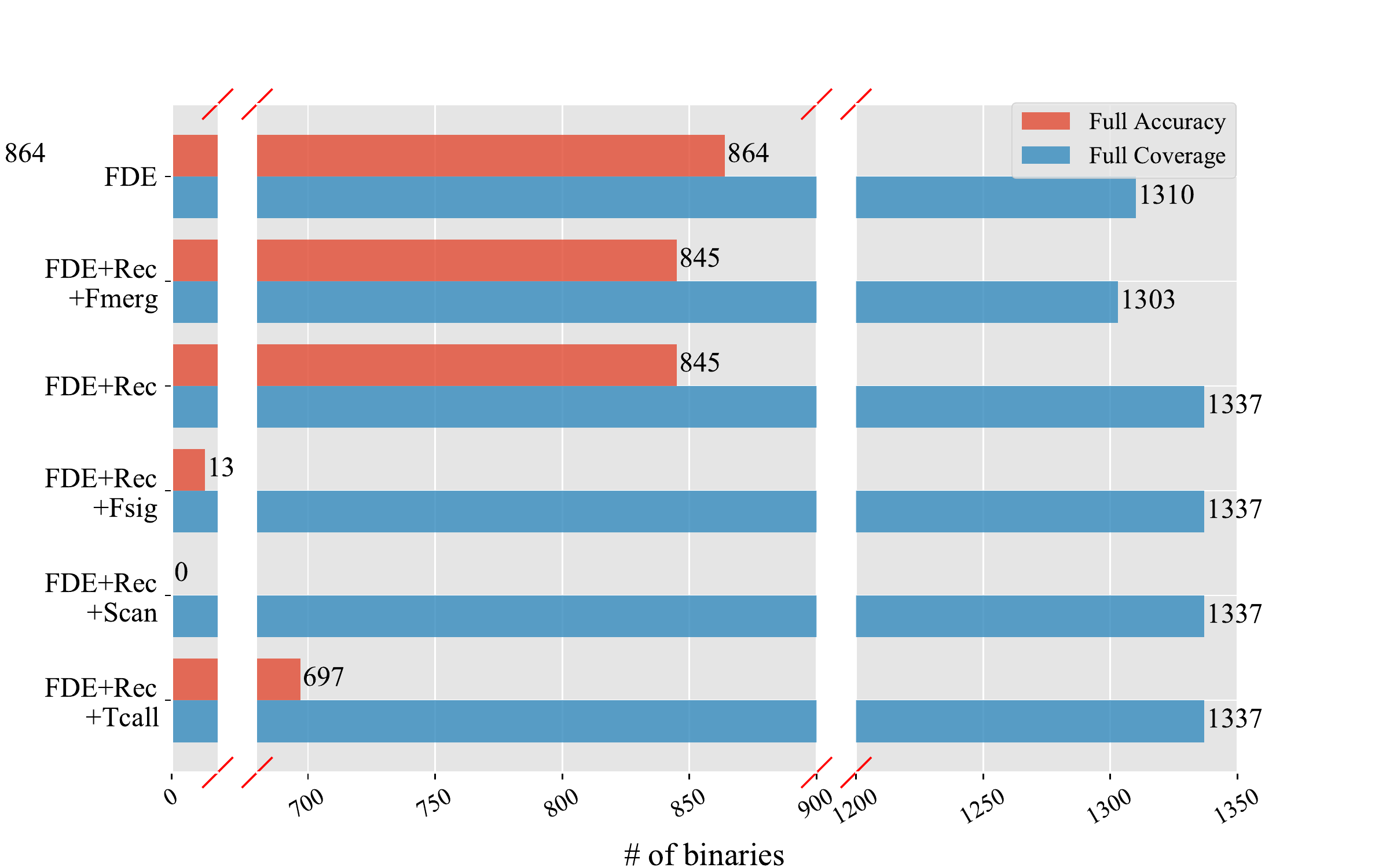}
    \caption{\angr (1,343 binaries)}
    \label{fig:full_cov_acc_angr}
  \end{subfigure}
  \begin{subfigure}[t]{0.32\textwidth}
  \centering
  \includegraphics[scale=0.26]{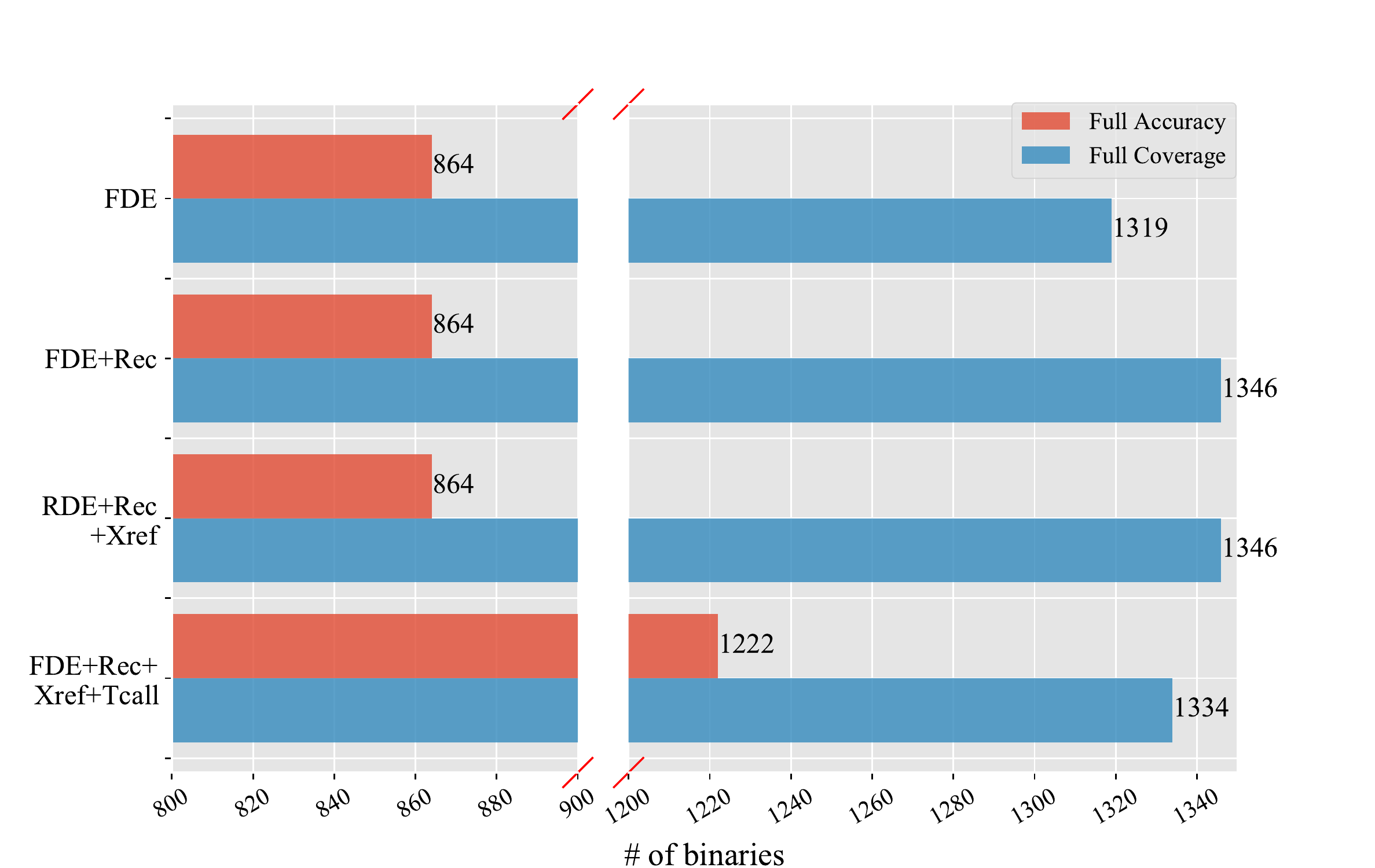}
    \caption{Optimal Strategies (1,352 binaries)}
    \label{fig:full_cov_acc_fetch}
  \end{subfigure}
  \caption{The number of binaries where different strategies achieve full coverage or full accuracy. In the figure, {\bf FDE} indicates solely using FDEs; {\bf Rec},  {\bf Fsig}, {\bf Tcall} respectively indicate running recursive disassembly, function matching, and tail call detection in the corresponding tool; {\bf CFR} indicates control flow repairing in \ghidra; {\bf Fmerg} and {\bf Scan} indicate function merging and linear scan by \angr; {\bf Xref} indicates function pointer detection in the optimal strategies.}
  \label{fig:full_cov_acc}
  \vspace{-0.8em}
\end{figure*}

% \begin{figure}[t]
%     \centering
%     \includegraphics[scale=0.42]{figure/full_cov_acc_ghidra.pdf}
%     \caption{The number of binaries with full coverage and full accuracy in ghidra's different configurations. \textbf{\small ES} - Functions that present in eh\_frame and symbols. \textbf{\small R} - Recursively disassembling from known functions. \textbf{\small RP} - Enable control flow repairing of ghidra. \textbf{\small FM} - Enable function matching.}
%     \label{fig:full_cov_acc_ghidra}
%     \vspace{-1.2em}
% \end{figure}

% \begin{figure}[t]
%     \centering
%     \includegraphics[scale=0.42]{figure/full_cov_acc_fetch.pdf}
%     \caption{The number of binaries with full coverage and full accuracy in FETCH's different configurations. \textbf{\small ES} - Functions that present in eh\_frame and symbols. \textbf{\small R} - Recursively disassembling from known functions. \textbf{\small XR} - Disassembling based on x-ref. \textbf{\small T} - Enable tailcall detection.}
%     \label{fig:full_cov_acc_fetch}
%     \vspace{-1.2em}
% \end{figure}

\myparatight{Results with \ghidra} \ghidra can run all the 1,352 self-built binaries. However, the results are well below expectations. In comparison to solely using FDEs, recursive disassembly by \ghidra significantly reduced the coverage: the total number of covered function starts dropped from 1,103,832 to 1,088,377; and the number of binaries with non-detected functions increased from 33 to 78. The reduction of coverage is mainly caused by a strategy --- \emph{control-flow repairing} --- that examines the function start after a non-returning function. If the function start cannot be reached by other control flows, \ghidra removes that function start. Due to inaccuracy in the detection of non-returning functions and incompleteness in the analysis of control flows~\cite{sok:sp20}, control-flow repairing often removes many true function starts, leading to reduced coverage as we observed.

We then conducted a follow-up test where we disabled the control-flow repairing. This time \ghidra's recursive disassembly demonstrated its effectiveness: in comparison to solely using FDEs, it increased the number of detected functions from 1,103,832 to 1,104,786 and dropped the the number of binaries with non-detected functions from 33 to only 6 (see Figure~\ref{fig:full_cov_acc_ghidra}). While the recursive disassembly by \ghidra indeed brings more coverage, we found that it is accompanied by another heuristic to detect thunk functions. The heuristic considers a function that starts with a jump to be a thunk function and takes the target of the jump as a new function start. In our test, the heuristic introduced over 400 new false positives and increased the number of binaries that have false positives from 488 to 542.

\myparatight{Results with \angr} \angr can only run 1,343 of the self-built binaries because it could not open the remaining 9. Before discussing the results, we want to note an observation that can make a significant difference. Specifically, \angr marks a special group of functions (or precisely, functions that have a basic block solely consisting of padding instructions) as \emph{alignment}. A recent study~\cite{sok:sp20} suggests excluding the alignment functions for comparison. However, we found that doing so will reduce the coverage but not improve accuracy. Therefore, we preserved all the alignment functions.

In total, the group of 1,343 binaries contain 982,763 functions. By using FDEs alone, we detected 981,317 of those functions and achieved full coverage in 1310 binaries. However, the further recursive disassembly by \angr decreased the number of binaries with full coverage to 1,303. The major cause is a heuristic that \angr uses to merge functions. To be specific, \angr merges two adjacent functions if the two functions are connected by a jump which is the only outgoing control-transfer from the first function and the only incoming control-transfer to the next function.  

Following up the above test, we re-ran \angr's recursive disassembly without function merging. This time recursive disassembly demonstrated true effectiveness. It increased the number of detected functions from 981,317 to 982,195 and increased the number of binaries with full coverage from 1,310 to 1,337. However, similar to \ghidra, \angr's recursive disassembly is coupled with a heuristic that introduces extra false positives. In an alignment function where the beginning instructions are considered padding, the heuristic will mark the first non-padding instruction a new function start, incurring 3,973 false positives.

%Angr will merge functions if following condition meet:
%	The second function is not called by other code.
%	The first function has only one jumpout site, which points to the second function
%The first function and the second function are adjacent.

%# of full coverage: 1286
%# of full precision: 845
%# of functions: 982763
%# of total detected functions: 981811

%# of binaries: 1343+ (9 could not open)
%# of full coverage: 1239

\myparatight{Results with Safe Recursive Disassembly} As described above, the recursive disassembly by \ghidra and \angr is coupled with heuristics. In addition, the recursive disassembly itself in \ghidra and \angr also uses other unsafe strategies to handle complex constructs (\eg indirect jumps)~\cite{sok:sp20}. These indicate that the tests with \ghidra and \angr do not truly unveil the coverage of "safe" recursive disassembly on top of FDEs. This motivated us to run an extra test with error-free recursive disassembly. In general, recursive disassembly can run into errors only when handling \emph{indirect jumps}, \emph{indirect calls}, \emph{tail calls}, and \emph{non-returning functions}. We handle these complex constructs as follows to avoid errors.

\begin{itemize}[leftmargin=*]
\item[\ding{182}] \emph{Indirect Jumps} -  We only consider indirect jumps for jump tables. Specifically, we follow \dyninst~\cite{meng2016binary} to detect and solve jump tables which has proven high precision~\cite{sok:sp20} and fixed some implementation defects in \dyninst. 
% Given an indirect jump, we run backward slicing with the target as the source. In the sliced area, if the first memory read has the format of {\tt [CONST + reg * size]}, we consider the indirect jump as a jump table which uses {\tt CONST} and {\tt reg} as the base address and index, respectively. 
% From the index, we run backwards slicing again until the function start. 
% Then, we re-use \dyninst's intra-procedure Value Set Analysis (VSA) to determine the value bounds of the index. As shown by~\cite{sok:sp20}, this algorithm can effectively resolve jump tables with high precision. 
% The only errors are caused by defects in \dyninst's implementation and we fixed them accordingly. 

\item[\ding{183}] \emph{Indirect Calls} -  We skip all indirect calls. 

\item[\ding{184}] \emph{Tail Calls} -  We do not detect tail calls.

\item[\ding{185}] \emph{Non-returning Functions} -  We reuse \dyninst's algorithm to detect non-returning functions, which has proven accurate~\cite{sok:sp20}. 
% One thing we want to note is that \dyninst uses an incomplete list of non-returning library functions, often leading to missing cases. 
We expanded the non-returning library functions used by \dyninst to cover all the cases in our self-built binaries. 
In particular, we handled {\tt error} and {\tt error\_at\_line} as special cases since they are non-return only when the first argument is non-zero.
% In the list, {\tt error} and {\tt error\_at\_line} are two special cases. They are non-returning only when the first argument is non-zero. 
Encountering either function, we run backward slices from the first argument and examine whether the argument always flows from 0. If so, we consider the function returning and non-returning otherwise.
\end{itemize}

Running the above error-free recursive disassembly, we achieved identical coverage as \ghidra and \angr, while more importantly, we introduced no false positives during the recursive disassembly, as illustrated by Figure~\ref{fig:full_cov_acc}.

%\vspace{1ex}
%\begin{mdframed}
%\noindent\textbf{Summary:} Recursive disassembly on top of FDEs can  %benefit the coverage of function starts. To maximize the benefit, it is %sufficient to only run the conservative, heuristic-free version of %recursive disassembly. Involving further heuristics from existing tools %(\ghidra and \angr) can hurt the accuracy while bringing no extra coverage %or even reducing the coverage.
%\end{mdframed}

\subsection{Answering Question $\mathbb{Q}_3$} After running our safe recursive disassembly with FDEs, we missed 568 functions in the 1,352 binaries and missed 568 functions in the 1,343 binaries that \angr can run. All the missed functions are assembly functions, mainly belonging to two groups: (1) functions that are only reachable via tail calls and the successors of those functions (2) functions that are only reachable via indirect calls and the successors of those functions. In the last study, we aim to understand whether the other unsafe approaches used by \ghidra and \angr can help detect those functions and what harm they will incur.

\myparatight{Results with \ghidra} \ghidra uses two heuristic-based approaches, function matching and tail call detection,\footnote{Tail call detection is not enabled by default in \ghidra (neither in \angr). We tested it because some missing function starts are due to tail calls.} to further detect function starts. We tested the two approaches in turn. As indicated by Figure~\ref{fig:full_cov_acc_ghidra} the two approaches are not helpful to coverage. The function matching detected no new function starts, despite it neither brought false positives. The tail call detection found 16 new function starts, however, at the cost of 97,339 new false positives. 

\myparatight{Results with \angr} Besides using function matching and tail call detection, \angr also detects function starts in its linear-scan process. In our study, we in turn tested function matching, tail call detection, and linear scan. Function matching helped detect 8 new function starts. However, it brought 4,128 false positives and it decreased the number of binaries with full accuracy from 845 to 13; The tail call detection found 211 new function starts at the cost of 4,686 false positives. Moreover, it dropped the number of binaries with full accuracy from 845 to 697; The linear scan detected 230 new function starts. However, it increased the number of false positives from 35,159 to 210,921 and more importantly, it eliminated all the binaries that have full accuracy.

\subsection{Moving Towards Full Coverage} 
\label{subsec:xref-coverage}
Recursive disassembly with FDEs can provide high coverage, but it does not guarantee full coverage for a specific binary. It would be very convenient if we could identify --- or slightly over-approximate --- the missing function starts, since this will enable the users to have full coverage and, with slight manual efforts on examining the functions we further identify, obtain a full accuracy (except the accuracy issues inherited from FDEs). To this regard, we explored another soundness-driven approach~\cite{qiao2017function} and we present the details as follows. 

Given a binary, we first run recursive disassembly on top of FDEs and then collect all the potential function pointers. For each pointer, we validate its legitimacy. Specifically, we run our conservative recursive disassembly from the pointer and check four types of errors: (i) invalid opcodes; (ii) running into the middle of previously disassembled instructions; (iii) control transfers to the middle of previously detected functions; and (iv) invalid calling conventions (to validate calling conventions, we use the rule that all non-argument registers, namely registers other than rdi, rsi, rdx, rcx, r8, r9, must be initialized before use). If no error occurs, we consider the function pointer legitimate and take it as a new function start. 

A key challenge in the above approach is the identification of function pointers. To overcome this challenge, we take a conservative approach to collecting a super-set of function pointers. Technically, we scan every consecutive eight-bytes (\eg [0,...,7], [1,...,8], [2,...,9], \etc) in the data segment and the non-disassembled regions, considering each of the eight-bytes as a pointer. We also identify all the constant operands in the disassembled code and consider each constant as a potential pointer. As demonstrated by a recent study~\cite{sok:sp20}, this combined strategy can collect all potential function pointers. We further want to note that, once we determine a ``legitimate'' function pointer, we will update the pointer collection based on the results of recursive disassembly from that pointer.  

Applying the above approach to our 1,352 self-built binaries, we detected 154 more function starts without introducing new false positives. %On average, our approach reported 0.31 missing function starts for each binary, which requires minor effort to validate. 
We also examined the 414 functions we still missed. These functions belong to two categories. The first category includes 160 unreachable assembly functions (\ie assembly functions that are not referenced anywhere and their successors). Missing such functions is in principle harmless. The second category contains 254 functions that are only referenced by tail calls in the same function. As we will explain in \S~\ref{subsec:fixerrors}, the side effect of missing the 254 functions is equivalent to in-lining them into their parent functions, which is also in general harmless.  %all the 254 functions can be detected by our new tail call detection algorithm (see \S~\ref{subsec:fixerrors}).
We finally want to note that while our function pointer detection is not theoretically safe, on average it only reports 0.31 function starts for each binary, which requires minor human efforts to validate.

\section{Improving Accuracy of Function Start Detection with Call Frames}
\label{sec:study-accuracy}

In this section, we aim at our second research goal --- \emph{understanding the accuracy of function start detection with exception handling information}. Existing tools, \ghidra and \angr, simply trust the fidelity of FDEs when using them for function detection. However, we found that FDEs are not perfectly accurate and in fact, they can introduce many false positives. In the following, we will first unveil and quantify the false positives that FDEs can introduce, and then we will present a new approach to fix the errors. 

\begin{figure}[!t]
  \centering
\begin{subfigure}{0.23\textwidth}
\begin{lstlisting}[language={[x86masm]Assembler},basicstyle={\tiny\ttfamily},numberstyle=\tiny,deletekeywords={df},xleftmargin=5mm]
; start of part 1, FDE1
4126c0: push   rbp
...
4128ff: test   rax,rax
; a non-tail-call jump to part 2
412902: je     404fbe 
; end of part 1
; gap
; ...
; start part 2, FDE2
404fbe: mov    esi,$0x4437e0
404fc3: xor    edi,edi
...
405010:  call  4247d0 
405015:  jmp    404fdc 
; end of part 2
\end{lstlisting}
  \caption{A non-contiguous function from BinUtils-2.26. The function contains two parts and each part has a separate FDE.}
  \label{fig:nc-func}
\end{subfigure}
~\begin{subfigure}{0.23\textwidth}
\begin{lstlisting}[language={[x86masm]Assembler},basicstyle={\tiny\ttfamily},numberstyle=\tiny,deletekeywords={df},morekeywords={PC,Begin,Range,CFIs,syscall},xleftmargin=5mm]
; start of function
3c610 <__restore_rt>:
  3c610: mov    $0xf,%rax
  3c617: syscall
  3c619:
    
; FDE Entry
00021e98 FDE
  PC Begin: 3c60f
  PC Range: a
  CFIs:
     DW_CFA_expression: reg8 DW_OP_breg7 +40
     DW_CFA_expression: reg9 DW_OP_breg7 +48
     DW_CFA_expression: reg10 DW_OP_breg7 +56
     ...
     DW_CFA_nop:
\end{lstlisting}
  \caption{A handwritten function in Glibc-2.27. The begin of handwritten FDE does not equal to the begin of function start.}
  \label{fig:handwritten-fde}
\end{subfigure}
\vspace{0.5em}
\caption{Examples of false positive introduced by FDEs.}
  \vspace{-1.0em}
\end{figure}

\subsection{Identifying and Quantifying Errors Due to FDEs}
\label{subsec:identifyerrors}

To systematically understand the errors introduced by FDEs, we compared the function starts extracted from FDEs in our self-built binaries with the ground truth. We found that FDEs brought 34,772 false positives, spanning 488 of the 1,352 binaries. In the case of Mysqld compiled with GCC and Ofast, FDEs introduced 3, 616 false positives. 

Among all the false positive, 34,769 are related to non-contiguous functions. For each non-contiguous function, the compiler inserts separate FDEs for different parts in the function  (\eg Figure~\ref{fig:nc-func}). As such, function starts extracted from FDEs for the non-beginning parts become false positives. Such false positives are rooted from the design of FDEs: a single FDE cannot cover multiple non-contiguous code segments. Without adapting the design and the standard behind, it is unlikely to fully avoid such false positives. {\em We also want to note that symbols have the same problem: separate symbols are generated for different parts from the same non-contiguous function. Our study shows that symbols also introduce the 34,769 false positives.} The remaining 3 FDE false positives are from assembly functions where the developers (intentionally) insert CFI directives that label the incorrect function starts (\eg Figure~\ref{fig:handwritten-fde}).
% \url{https://github.molgen.mpg.de/git-mirror/glibc/blob/master/sysdeps/unix/sysv/linux/x86_64/sigaction.c#L146}). 

% \vspace{-2.5em}
The false positives brought by FDEs can hurt security applications. For instance, many control flow integrity solutions for binary code~\cite{van2016tough,zhang2013control,erlingsson2006xfi,muntean2018tau} consider all function starts as legitimate targets of indirect control transfers. Our experiment with ROPgadget~\cite{shellsto38:online} shows that the basic blocks at the FDE-introduced false function starts contain 99,932 valid ROP gadgets. Including the FDE-introduced false function starts would make control hijacks to those ROP gadgets un-detectable, reducing the security effectiveness of control flow integrity. However, we observe that both \ghidra and \angr do not provide any solution to address the false positives introduced by FDEs: all the the false positives due to FDE persist across every detection step of \ghidra and \angr.

%we tracked all the false positives due to FDE in the analysis-cycle of \ghidra and \angr, and found that all the false positives persist across every detection step. 
% \vspace{-1.6em}
\subsection{Fixing Errors Due to FDEs} 
\label{subsec:fixerrors}

% \vspace{-6.5em}

\begin{algorithm}[t]
\caption{Tail-call Detection and Function Merging}
\label{alg:tail-call}

\begin{algorithmic}[1]
\scriptsize
  \State \emph{Input}: A list of functions $L$
\Function{TailCall-Detect}{$L$}
\For {$f \in L$}
  \For {direct/conditional jump $ j\in f$}
  \State $t \gets$ Target($j$)
  \State isTailCall = False
  \If {$t \in f$}
    \State continue \algorithmiccomment{Skip jump inside function}
  \EndIf
  \State $h \gets$ getStackHeight($j$)
  \If {$h = 0$}
    \If {HasRefTo($t$, $L$) $\land$ MeetCallConv($t$)}
    \State add\_tail($t$) \algorithmiccomment{Find a tail call}
    \State add\_func($L$, $t$)
    \State isTailCall = True
    \EndIf
  \EndIf
  
  \If {$\neg$isTailCall $\land$ IsFunction($t$) $\land$ RefTo($t$) == $j$}
    \State MergeFunc($t$, $f$) \algorithmiccomment{Merge function}
    \State remove\_func($L$, $t$)
  \EndIf
  
  \EndFor
\EndFor
\State \Return $L$
\EndFunction
\end{algorithmic}
\end{algorithm}
\setlength{\textfloatsep}{1.2em}

% \subsubsection{Algorithm Design}
To effectively reduce the FDE-introduced false positives, we propose a new algorithm based on a key observation that \emph{two distant parts in the same non-contiguous function are connected via a jump}. In principle, by determining that the jump is not a jump between two functions (\ie not a tail call), we can safely merge the two distant parts. In general, it is hard to design a perfect algorithm to detect tail calls. Most existing tools use heuristics that are neither sound nor complete~\cite{sok:sp20}. In this work, we propose a completeness-driven, but safe, algorithm, which ensures fidelity of the captured cases and minimizes the side effects of the missed cases. 

Our algorithm is shown in Algorithm~\ref{alg:tail-call}. It iterates over each conditional or unconditional jump in each function. It considers a jump to be a tail call if:

\begin{itemize}[leftmargin=*]
\setlength\itemsep{0em}
\item[\ding{182}] The stack pointer at the jump site is right below the return address. This is a must-be-true property of tail call because the current function has to ensure that the target can directly return to its parent function.

\item[\ding{183}] The target satisfies the calling conventions since the target of a tail call must be a new function. To validate calling conventions, we re-use the rule as described in \S~\ref{subsec:xref-coverage}.

\item[\ding{184}] The target is not referenced elsewhere other than jumps in the current function. In theory, a tail call does not have to meet this rule. However, our empirical studies with a large-corpus binaries (listed in Table~\ref{tab:dataset}) show that this rule can perfectly avoid false positives. More importantly, this rule ensures that the target of any missed tail call is not referenced elsewhere. Thus, the side effect of the missed tail call is equivalent to in-lining the target function to the source function, which should be generally harmless. 

\end{itemize}
For a jump that we determine to be not a tail call, we check whether the target has an FDE record and whether the target is not referenced elsewhere. If both conditions hold, we consider the jump part and the target part are from the same function and we merge the two parts to the same function. 

To implement Algorithm~\ref{alg:tail-call}, there are two challenges. First, it needs to know the value of the stack pointer at a jump site. Many tools, such as \dyninst~\cite{meng2016binary} and \angr~\cite{shoshitaishvili2016sok}, include static analysis of stack height. However, as shown in Table~\ref{tab:stackheight_result}, these analyses can often provide inaccurate stack height due to side effects of other errors and defects of engineering. To address this challenge, we opt to use the stack height recorded by CFIs in FDEs. For conservativeness, we only pick functions whose CFIs give complete information of stack height, by checking (i) whether the CFA in the CFIs is represented by {\tt rsp} and the CFA is initialized as {\tt rsp+8} and (ii) whether a {\tt DW\_CFA\_def\_cfa\_offset} instruction exists wherever the stack height is changed. We skip functions with incomplete stack height. Second, the algorithm needs to collect all the references to functions. We overcome this challenge by using the conservative approach in \S~\ref{subsec:xref-coverage}.

We also looked at the 3 false positives introduced by the developers. We found that the code blocks pointed to by those FDEs all present invalid calling conventions (using the rules in \S~\ref{subsec:recdias}). By checking the calling conventions of each function directly identified from FDEs, we detected the three false positives. After removing the false positives and re-running our pointer-based detection in \S~\ref{subsec:xref-coverage}, we also identified the false negatives masked by those three false positives.

\begin{table*}[t!]
\setlength\tabcolsep{1pt}
  \centering
  \setlength\tabcolsep{5.3pt}
  \caption{Comparison results between \tool and existing tools. The results highlighted in blue indicate the best results. FP \#: the number of false positives ({\em thousands}); FN \#: the number of false negatives ({\em thousands}).} 
  \label{tab:funcentryresult}
  \scriptsize
  \begin{tabular}{l|cc|cc|cc|cc||cc|cc||cc|cc||cc}
    \hline\hline
    \multirow{2}{*}{\textbf{OPT}}  & \multicolumn{2}{c|}{\bf \dyninst}  & \multicolumn{2}{c|}{\bf \bap} & \multicolumn{2}{c|}{\bf \radare}  & \multicolumn{2}{c||}{\bf \nucleus} &
    \multicolumn{2}{c|}{\bf \ida} & \multicolumn{2}{c||}{\bf \ninja}  &
    \multicolumn{2}{c|}{\bf \ghidra} &
    \multicolumn{2}{c||}{\bf \angr} &
    \multicolumn{2}{c}{\bf \tool} \\\cline{2-19}
     & FP \# & FN \# & FP \# & FN \# & FP \# & FN \# & FP \# & FN \# & FP \# & FN \# & FP \# & FN \# & FP \# & FN \# & FP \# & FN \# & FP \# & FN \# \\ \hline
    O2  &  12.20 & 81.41  & 148.94  &  93.82 &  4.10 & 100.23  &  17.49 &  18.41 & 2.68 &  37.04 & 41.17  & 11.96 & 43.49 & 5.62 & 51.68 & 0.20 & \textcolor{blue}{0.78}  & \textcolor{blue}{0.08}  \\
    O3 &  12.60 & 82.05  &  165.06 &  96.71 &  4.26 & 106.99  & 20.15  &  16.70 & 2.71  &  36.94 & 44.57  &  12.80 & 46.62 & 4.65 & 54.71 & 0.19 & \textcolor{blue}{0.84}  & \textcolor{blue}{0.08}  \\
    Os &  6.72 &  87.82  &  109.50 & 79.86  & 3.08  & 81.67 &  23.35 &  27.86 & 0.97 & 32.74  & 34.45  & 6.14  & 0.92 & 1.97 & 37.10 & 0.18  & \textcolor{blue}{0.07}  & \textcolor{blue}{0.14}  \\
    Of & 13.63 & 88.22  & 159.41 &  92.20 & 3.08  & 93.93  & 26.68  & 19.36  & \textcolor{blue}{0.86}  & 37.97  &  40.08 & 10.39 & 46.43 & 4.66 & 67.43 & 0.18 & 0.97  & \textcolor{blue}{0.12}  \\ \hline
    {\bf Avg.}  &  {\bf 11.29} & {\bf 84.88}  & {\bf 132.48}  & {\bf 90.65}  & {\bf 3.63}  & {\bf 95.71}  & {\bf 21.92}  & {\bf 20.58}  & {\bf 1.81} & {\bf 36.17}  & {\bf 40.07}  & {\bf 10.32} & {\bf 34.37} & {\bf 5.23} & {\bf 52.73} & {\bf 0.19}   & \textcolor{blue}{\bf 0.67}  & \textcolor{blue}{\bf 0.11} \\\hline\hline
  \end{tabular}
  \vspace{-0.8em}
\end{table*}

% \subsubsection{Algorithm Evaluation}
\subsection{Algorithm Evaluation}

We tested the performance of Algorithm~\ref{alg:tail-call} with our 1,352 self-built binaries. On top of FDEs, we first ran our recursive disassembly and our function pointer detection. Then, we ran Algorithm~\ref{alg:tail-call} and measured the change of both coverage and accuracy. In total, our algorithm reduced the number of FDE-introduced false positives from 34,772 to 2,659, increasing the number of binaries with full accuracy from 864 to 1,222. 

Among the remaining 2,659 false positives, 2,656 are still caused by non-contiguous functions. Our algorithm missed detecting them because CFIs in those functions do not provide complete stack height information, and thus, we skipped processing those functions. While intuition suggests we can re-use static analyses from existing tools (\eg \angr and \dyninst) for stack height information in such functions, we opted not to do so. The reason, as aforementioned, is that the static analyses can be incomplete or inaccurate. To validate our choice, we also conducted an empirical evaluation. Specifically, we compared the stack height information from CFIs and the stack height information provided by both \angr and \dyninst. It is worth noting that we only ran the comparison on functions whose CFIs provide complete stack height information. As shown by the results in Table~\ref{tab:stackheight_result}, the stack height analyses by \angr and \dyninst carries both incompleteness and inaccuracy (even just considering the jump sites), using of which can hurt our tail call detection.

\begin{table}[t!]
  \centering
  \caption{Coverage and precision of stack height analyses by \angr and \dyninst. Baseline is the stack height from CFIs. {\bf Full} indicates the result with all code locations and {\bf Jump} indicates the result with only jump sites considered.}
  \label{tab:stackheight_result}
  \scriptsize
  \begin{tabular}{c|cc|cc|cc|cc}
    \hline\hline
    \multirow{3}{*}{\textbf{OPT}}  & \multicolumn{4}{c|}{\bf \angr} & \multicolumn{4}{c}{\bf \dyninst} \\\cline{2-9}
    & \multicolumn{2}{c|}{\bf Full} & \multicolumn{2}{c|}{\bf Jump}
    & \multicolumn{2}{c|}{\bf Full} & \multicolumn{2}{c}{\bf Jump} \\\cline{2-9}
     & Pre & Rec & Pre & Rec
     & Pre & Rec & Pre & Rec
     \\ \hline
 %   O0 & 99.98 & 99.99 & 95.40 & 99.07 & 99.92 &  97.69 &  99.94  & 99.99  &  97.14  & 91.77  & 98.92  & 97.17  &  99.96 &  97.33 & 98.82  & 99.84  & 99.30  &  98.81 & 99.63  & 99.54  & 99.99  & 99.99  \\
 %   O1 & 99.88 & 99.99 & 86.43 & 93.08 &  93.51 & 60.62  & 99.84  &  99.96 & 79.84   &  67.79 & 98.67  &  59.60 & 99.70  & 79.38  & 98.69  &  98.11 &   &   & 97.35  &  94.93 & 99.09  & 99.99  \\
    O2 & 93.00 & 97.26 & 98.56 & 95.95 & 93.18 & 98.26 & 98.60 & 99.36 \\
   O3 & 93.66 & 97.28  & 98.62 & 95.84 & 92.87 & 97.96 & 98.50 & 99.34  \\
    Os & 96.42 & 99.22 & 99.27 & 99.27 & 99.10 & 98.27 & 98.67 & 99.35 \\
    Of & 93.20 & 97.08 & 98.43 & 94.53 & 94.08 & 98.60 & 98.90 & 99.36 \\ \hline
     {\bf Avg.} & {\bf 94.07} & {\bf 97.71}  & {\bf 98.72} & {\bf 96.40} & {\bf 94.81} & {\bf 98.27} & {\bf 98.67} & {\bf 99.35} \\\hline\hline
  \end{tabular}
%   \vspace{-1.8em}
\end{table}

We finally examined whether Algorithm~\ref{alg:tail-call} brought false negatives and false positives. It turns out that the algorithm did not bring extra false positives, but it introduced 161 new false negatives, slightly reducing the number of binaries with full coverage from 1,346 to 1,334 (see Figure~\ref{fig:full_cov_acc_fetch}). All the 161 false negatives are because we merged targets of true tail calls to the call sites. Despite missing the 161 functions slightly affects coverage, the 161 functions are only referenced by tail calls in a single function (otherwise they will be detected by our algorithm). In this sense, the side effect of missing those functions is equivalent to in-lining them to their parent functions, which in general produces no harm. This is also the reason why the 254 false negatives we discussed in \S~\ref{subsec:xref-coverage} are harmless. 

%\begin{mdframed}
%\noindent\textbf{Summary:} FDEs may not point to function starts, and %therefore, can introduce a significant number of false positives in %function detection. Existing tools that use FDEs do not seek to detect and %fix those false positives. We propose a new algorithm which can %effectively reduce the false positives, making FDEs more reliable.
%\end{mdframed}

%34769 -> 2656
% 864 -> 1222

%remaining: 2656

%

\section{Comparing with Other Approaches}
\label{sec:eval}

We finally conducted an extra comparison, where we compared the optimal strategies of using FDEs and 8 existing tools (using the \binnum self-built binaries shown in Table~\ref{tab:dataset}). For simplicity of presentation, we will use {\bf \tool} (\ul{\bf F}unction d\ul{\bf ET}ection with ex\ul{\bf C}eption \ul{\bf H}andling) to represent our optimal strategies of using FDEs.
%In the rest of this section, we will first describe the setup of \tool and the 7 existing tools and then present the results of recall and precision.

%combing FDEs with our recursive disassembly, our function pointer detection, and our tail call detection and (2) 7 existing approaches that do not use FDEs. In the following, we will first descr

%FETCH: FUNCTION DETECTION WITH EXCEPTION HANDLING

\myparatight{Setup} \tool works by first extracting FDEs and then running our safe recursive disassembly (\S~\ref{subsec:recursive}), our function pointer detection (\S~\ref{subsec:xref-coverage}), and our tail call detection (\S~\ref{subsec:fixerrors}). The 6 other tools that do not use FDEs are from two categories: (i) open-source tools that are designated for function detection or have a component of function detection (\dyninst~\cite{meng2016binary}, \bap~\cite{brumley2011bap}, \radare~\cite{radare2_org}, and \nucleus~\cite{andriesse2017compiler}, and (ii) commercial tools that can detect functions (\ida~\cite{eagle2011ida} and \ninja~\cite{binaryni92:online}). Their configures are same as~\cite{sok:sp20}. The results of \ghidra and \angr 
are also included for convenience of comparison.

% \begin{itemize}[leftmargin=*]
% %\setlength\itemsep{0em}
% %\item[\ding{182}] {\bf \angr -} We ran \angr with its {\em CFGFast} interface for recursive disassembly. We set {\em detect\_tail\_calls} to enable tail call detection. We also excluded functions marked as ``alignment" when collecting its results. 

% \item[\ding{182}] {\em \dyninst -} We ran \dyninst with its {\em ParseAPI} to do recursive disassembly and we set the {\em IdiomMatching} option to enable the decision-tree based function detection. 

% %\item{\bf \ghidra -} We ran \ghidra with default settings plus the {\em Assume Contiguous Functions Only} and {\em Allow Conditional Jumps} options to enable tail call detection.

% \item[\ding{183}] {\em \bap -} We ran \bap with {\em -dasm} and {\em -drcfg} to do disassembly and reconstruct the CFG. We used the {\em with-no-return} ~\cite{bap_noreturn} pass to detect non-returning functions.

% \item[\ding{184}] {\em \radare -} We ran \radare with three options: {\em aa} for 
% default recursive disassembly, {\em aanr} for non-return function detection, and {\em aap} for signature-based function matching. 

% \item[\ding{185}] {\em \nucleus -} We ran \nucleus as it is.

% \item[\ding{186}] {\em \ida, \ninja -} We ran their default settings.

% \end{itemize}
% \vspace{-0.5em}

\myparatight{Coverage and Accuracy} As shown in Table~\ref{tab:funcentryresult}, \tool presents extremely high coverage and accuracy. It only brings hundreds (or dozens) false positives and false negatives from the total \binnum binaries, regardless of the optimization level. \tool outperforms all the 8 other tools. It produces the best coverage in all the settings and the best accuracy except under optimization level Ofast. These results demonstrate the benefits of the FDE-assisted solutions.

%\tool also demonstrates unprecedented advantages. It outperforms all other tools in terms of recall and it has a precision very close to the best results produced by \ida (the flagship commercial tool of reverse engineering). Most importantly, \tool presents an overall better result than symbols: \tool has a recall close to symbols (99.84\% \vs 99.99\%) while \tool has a considerably higher precision (99.96\% \vs 97.62\%). 

\begin{table}[t!]
  \centering
  \setlength\tabcolsep{1.0pt}
  \caption{Average time for tools to run a binary (second).}
  \label{tab:efficiency}
  \scriptsize
  \begin{tabular}{l |c|c|c|c|c|c|c|c|c}
    \hline\hline
      {\bf Tool} & \multicolumn{1}{c|}{\bf \textsc{DYNINST}} & \multicolumn{1}{c|}{\bf \bap} & \multicolumn{1}{c|}{\bf \radare}  &
      \multicolumn{1}{c|}{\bf \nucleus}  &
      \multicolumn{1}{c|}{\bf \textsc{ghidra}}  &
      \multicolumn{1}{c|}{\bf \angr}  &
      \multicolumn{1}{c|}{\bf \textsc{IDA}} & \multicolumn{1}{c|}{\bf \textsc{Ninja}\xspace} & \multicolumn{1}{c}{\bf \tool} \\ \hline
     {\bf Ave.}  & 2.8 &  114.2 & 34.9 & 3.1 & 40.4 & 78.5 & 10.3 & 20.4 & 3.3 \\\hline\hline
  \end{tabular}
  \vspace{-1.0em}
\end{table}

\myparatight{Efficiency} We also measured the average time required by each tool to run a binary, and we show the results in Table~\ref{tab:efficiency}. Overall, \tool can finish analyzing a binary in around 3.3 seconds, which represents a high efficiency. 

\section{Discussion}
\label{sec:discuss}

In this section, we discuss the limitations and future directions of our research. 
% \subsection{Threats to Reliability}
\subsection{Threats to Validity} In this research, we focus on exploring the best strategies to (i) achieve optimal coverage and accuracy of using FDEs for function start detection and (ii) minimize the risks to the reliability. There exists potential threats to the fidelity of our findings. First, we concluded that running safe recursive disassembly on top of FDEs can achieve extremely high coverage with guaranteed reliability. However, recursive disassembly in practice may not ensure safety due to complex constructs like indirect jumps and non-returning functions. To reduce this threat, we have adopted the most conservative strategies to handle them (\S~\ref{subsec:recdias}). Second, the reliability of our approach to fixing FDEs-introduced errors is threatened by the completeness of the algorithm of tail call detection. To mitigate this threat, we adopted three restrictive criteria to detect tail calls, which have empirically proven completeness (\S~\ref{subsec:fixerrors}). Finally, as we unveiled in \S~\ref{subsec:identifyerrors}, developers may manually insert or modify the contents of \ehf (intentionally or unintentionally) in a way that introduces errors. This is a threat to the accuracy and we currently cannot avoid the threat. However, such errors rarely happen in practice and therefore, we envision it would not raise major concerns. 

\subsection{Generality of Study} Our study focuses on x64 System-V binaries because such binaries are guaranteed by the ABI to have call frames. However, this does not mean our study cannot be applied to other types of binaries. In fact, the methods used in our study are architecture independent and can work with any types of binaries that have call-frame-similar data structures. We have already conducted preliminary studies on other types of binaries and confirmed the availability of a structure similar to call frames. In particular, we found that x86 System-V binaries also widely carry FDEs, covering nearly all the functions 
% (coverage with symbols as the baseline: O2 - 99.99\%, O3 - 99.99\%, Os - 99.99\%, and Ofast - 99.99\%)
. We also discovered that x64 PE binaries adopt an FDE-similar data structure to support exception handling~\cite{x64excep71:online}, which contains the starts and boundaries of functions. Our preliminary results show that at least 70\% of the functions are covered by this structure. In addition, the ABI of Arm architecture also has the similar structure to support exception handling~\cite{arm-excep:online}. As a future work, we plan to extend our study to cover other types of binaries.
% and explore the optimal use of exception handling information to detect functions in those binaries. 

%\noindent\textbf{CFA.} CFA is decided by compilers, which typically adopt one of the two approaches.
%If the function uses a frame pointer (\eg \texttt{rbp}), the compiler will designate the frame pointer for the CFA. Otherwise, the compiler designates the stack pointer (\eg \texttt{rsp}) for the CFA. In real-world binaries which are mostly compiled with higher optimization levels, the first option is barely used because frame pointers are typically omitted to avoid the extra preparation and clean-up. \cp{Another reason: has one more general register -- rbp.} Therefore, we will next explain how to actually do stack unwinding with CFIs that use the second option for CFA (\ie designate the stack pointer for CFA), still following the example in Fig~\ref{fig:eh-example}.

%\input{related}

\section{Conclusion}
\label{sec:conclusion}
%More recently, the research community discovers a new source --- the call frames in exception handling section --- to identify function starts. People find that to enable exception handling, compilers are mandated by the ABI to emit a special call frame section in x64 binaries, providing start information for functions wherever possible. Mainstream binary analysis tools, such as \ghidra and \angr, have already incorporated the information from call frames to facilitate function start detection. However, the use of call frames by existing tools has two critical problems. First, beyond using call frames, they further use many other heuristic-based approaches to improve the coverage. Those approaches can often bring errors, but it is unclear whether they can bring more coverage. Second, they fully trust the fidelity of call frames, without realizing and handling the errors that call frames can introduce.

In this paper, we focus on studying the use of call frames to detect function starts. We found that the use of call frames by existing tools has two common problems. First, beyond using call frames and safe approaches, existing tools also run additional unsafe approaches to detect function starts, seeking to improve the coverage. However, the unsafe approaches can often introduce errors and their capacity of improving coverage is unclear. Second, the existing tools fully trust the information from call frames, without recognizing that call frames can also introduce errors. To gain a deeper understanding of the two problems and hence, bring insights towards optimal strategies of using call frames for function start detection, we conducted two studies. In the first study, we measured the coverage and accuracy of function starts detected by different approaches that existing tools run on top of call frames. Our key finding is that combing safe recursive disassembly and call frames can already achieve the maximal coverage, and additionally including other unsafe approaches cannot benefit the coverage but can hurt the accuracy and reliability of the results. In the second study, we systematically unveiled and quantified the errors that call frames can introduce. We further presented the first approach that can effectively fix nearly all the errors without introducing side effects. 
% \cp{Ruotong, Please update here.}

%In this paper, we aim to explore the optimal strategies of using call frames for function start detection. First, we investigate the best way of using call frames to achieve optimal coverage with maximal reliability. Specifically, we study the coverage and accuracy of function starts detected by different approaches that existing tools run on top of call frames. We found that combing heuristic-less recursive disassembly and call frames can already maximize the coverage. Further running other heuristic-based approaches does not benefit coverage but can bring plenty of false positives. Second, we systematically unveil and quantify the errors that call frames can introduce. We further develop the first approach that can effectively fix nearly all the errors without introducing side effects. 

%\clearpage

\section*{Acknowledgments}
\label{sec:acks}

We would like to thank our shepherd Miklos Telek and the anonymous reviewers for their feedback. This project
was supported by the Office of Naval Research (Grant\#:
N00014-16-1-2261, N00014-17-1-2788, and N00014-17-1-2787) and NSF  (Grant\#: CNS-1948489). Any opinions, findings, and conclusions or recommendations expressed in this paper are those of the authors and
do not necessarily reflect the views of the funding agency.
\clearpage

% Generated by IEEEtranS.bst, version: 1.12 (2007/01/11)

%\appendix

%\input{appendix}


\begin{thebibliography}{10}
\providecommand{\url}[1]{#1}
\csname url@samestyle\endcsname
\providecommand{\newblock}{\relax}
\providecommand{\bibinfo}[2]{#2}
\providecommand{\BIBentrySTDinterwordspacing}{\spaceskip=0pt\relax}
\providecommand{\BIBentryALTinterwordstretchfactor}{4}
\providecommand{\BIBentryALTinterwordspacing}{\spaceskip=\fontdimen2\font plus
\BIBentryALTinterwordstretchfactor\fontdimen3\font minus
  \fontdimen4\font\relax}
\providecommand{\BIBforeignlanguage}[2]{{%
\expandafter\ifx\csname l@#1\endcsname\relax
\typeout{** WARNING: IEEEtranS.bst: No hyphenation pattern has been}%
\typeout{** loaded for the language `#1'. Using the pattern for}%
\typeout{** the default language instead.}%
\else
\language=\csname l@#1\endcsname
\fi
#2}}
\providecommand{\BIBdecl}{\relax}
\BIBdecl

\bibitem{shellsto38:online}
``Ropgadget,'' \url{http://shell-storm.org/project/ROPgadget/}, 2011.

\bibitem{ghidra_org}
N.~S. Agency, ``Ghidra,'' \url{https://www.nsa.gov/resources/everyone/ghidra/},
  2019.

\bibitem{alves2019function}
J.~Alves-Foss and J.~Song, ``Function boundary detection in stripped
  binaries,'' in \emph{Proceedings of the 35th Annual Computer Security
  Applications Conference}, 2019, pp. 84--96.

\bibitem{andriesse2017compiler}
D.~Andriesse, A.~Slowinska, and H.~Bos, ``Compiler-agnostic function detection
  in binaries,'' in \emph{2017 IEEE European Symposium on Security and Privacy
  (EuroS\&P)}.\hskip 1em plus 0.5em minus 0.4em\relax IEEE, 2017, pp. 177--189.

\bibitem{bao2014byteweight}
T.~Bao, J.~Burket, M.~Woo, R.~Turner, and D.~Brumley, ``$\{$BYTEWEIGHT$\}$:
  Learning to recognize functions in binary code,'' in \emph{23rd USENIX
  Security Symposium}, 2014, pp. 845--860.

\bibitem{bernat2011anywhere}
A.~R. Bernat and B.~P. Miller, ``Anywhere, any-time binary instrumentation,''
  in \emph{10th ACM SIGPLAN-SIGSOFT workshop on Program analysis for software
  tools}.\hskip 1em plus 0.5em minus 0.4em\relax ACM, 2011, pp. 9--16.

\bibitem{brumley2011bap}
D.~Brumley, I.~Jager, T.~Avgerinos, and E.~J. Schwartz, ``Bap: A binary
  analysis platform,'' in \emph{International Conference on Computer Aided
  Verification}.\hskip 1em plus 0.5em minus 0.4em\relax Springer, 2011, pp.
  463--469.

\bibitem{chen2015stackarmor}
X.~Chen, A.~Slowinska, D.~sse, H.~Bos, and C.~Giuffrida, ``Stackarmor:
  Comprehensive protection from stack-based memory error vulnerabilities for
  binaries.'' in \emph{NDSS}, 2015.

\bibitem{dwarf2010dwarf}
D.~D. I.~F. Committee \emph{et~al.}, ``Dwarf debugging information format,
  version 4,'' 2010.

\bibitem{davi2015isomeron}
L.~Davi, C.~Liebchen, A.-R. Sadeghi, K.~Z. Snow, and F.~Monrose, ``Isomeron:
  Code randomization resilient to (just-in-time) return-oriented programming.''
  in \emph{NDSS}, 2015.

\bibitem{gas:online}
G.~Developers, ``Cfi directives,''
  \url{https://sourceware.org/binutils/docs/as/CFI-directives.html#CFI-directives},
  2020.

\bibitem{arm-excep:online}
A.~Docs, ``Exception handling abi for the arm architecture - abi 2018q4
  documentation,''
  \url{https://developer.arm.com/documentation/ihi0038/latest/}, 3 2021.

\bibitem{x64excep71:online}
M.~Docs, ``x64 exception handling,''
  \url{https://docs.microsoft.com/en-us/cpp/build/exception-handling-x64?view=vs-2019}.

\bibitem{eagle2011ida}
C.~Eagle, \emph{The IDA pro book}.\hskip 1em plus 0.5em minus 0.4em\relax No
  Starch Press, 2011.

\bibitem{elsabagh2017strict}
M.~Elsabagh, D.~Fleck, and A.~Stavrou, ``Strict virtual call integrity checking
  for c++ binaries,'' in \emph{2017 ACM on Asia Conference on Computer and
  Communications Security}.\hskip 1em plus 0.5em minus 0.4em\relax ACM, 2017,
  pp. 140--154.

\bibitem{erlingsson2006xfi}
{\'U}.~Erlingsson, M.~Abadi, M.~Vrable, M.~Budiu, and G.~C. Necula, ``Xfi:
  Software guards for system address spaces,'' in \emph{7th USENIX Security
  Symposium}, 2006, pp. 75--88.

\bibitem{he2017no}
W.~He, S.~Das, W.~Zhang, and Y.~Liu, ``No-jump-into-basic-block: Enforce basic
  block cfi on the fly for real-world binaries,'' in \emph{54th Annual Design
  Automation Conference 2017}.\hskip 1em plus 0.5em minus 0.4em\relax ACM,
  2017, p.~23.

\bibitem{hiser2012ilr}
J.~Hiser, A.~Nguyen-Tuong, M.~Co, M.~Hall, and J.~W. Davidson, ``Ilr: Where'd
  my gadgets go?'' in \emph{2012 IEEE Symposium on Security and Privacy
  (SP)}.\hskip 1em plus 0.5em minus 0.4em\relax IEEE, 2012, pp. 571--585.

\bibitem{koo2018compiler}
H.~Koo, Y.~Chen, L.~Lu, V.~P. Kemerlis, and M.~Polychronakis,
  ``Compiler-assisted code randomization,'' in \emph{2018 IEEE Symposium on
  Security and Privacy (SP)}.\hskip 1em plus 0.5em minus 0.4em\relax IEEE,
  2018, pp. 461--477.

\bibitem{koo2016juggling}
H.~Koo and M.~Polychronakis, ``Juggling the gadgets: Binary-level code
  randomization using instruction displacement,'' in \emph{11th ACM on Asia
  Conference on Computer and Communications Security}.\hskip 1em plus 0.5em
  minus 0.4em\relax ACM, 2016, pp. 23--34.

\bibitem{li2006address}
L.~Li, J.~E. Just, and R.~Sekar, ``Address-space randomization for windows
  systems,'' in \emph{2006 22nd Annual Computer Security Applications
  Conference (ACSAC'06)}.\hskip 1em plus 0.5em minus 0.4em\relax IEEE, 2006,
  pp. 329--338.

\bibitem{liu2018alphadiff}
B.~Liu, W.~Huo, C.~Zhang, W.~Li, F.~Li, A.~Piao, and W.~Zou, ``$\alpha$diff:
  cross-version binary code similarity detection with dnn,'' in
  \emph{Proceedings of the 33rd ACM/IEEE International Conference on Automated
  Software Engineering}, 2018, pp. 667--678.

\bibitem{lu2018system}
H.~Lu, M.~Matz, J.~Hubicka, A.~Jaeger, and M.~Mitchell, ``System v application
  binary interface,'' \emph{AMD64 Architecture Processor Supplement}, 2018.

\bibitem{meng2016binary}
X.~Meng and B.~P. Miller, ``Binary code is not easy,'' in \emph{25th
  International Symposium on Software Testing and Analysis}.\hskip 1em plus
  0.5em minus 0.4em\relax ACM, 2016, pp. 24--35.

\bibitem{muntean2018tau}
P.~Muntean, M.~Fischer, G.~Tan, Z.~Lin, J.~Grossklags, and C.~Eckert,
  ``$\tau$cfi: Type-assisted control flow integrity for x86-64 binaries,'' in
  \emph{International Symposium on Research in Attacks, Intrusions, and
  Defenses}.\hskip 1em plus 0.5em minus 0.4em\relax Springer, 2018, pp.
  423--444.

\bibitem{binaryni92:online}
B.~Ninja, ``binary.ninja,'' \url{https://binary.ninja/}, 2019.

\bibitem{sok:sp20}
C.~Pang, R.~Yu, Y.~Chen, E.~Koskinen, G.~Portokalidis, B.~Mao, and J.~Xu,
  ``Sok: All you ever wanted to know about x86/x64 binary disassembly but were
  afraid to ask,'' in \emph{42nd IEEE Symposium on Security and Privacy (SP)},
  2021.

\bibitem{pappas2012smashing}
V.~Pappas, M.~Polychronakis, and A.~D. Keromytis, ``Smashing the gadgets:
  Hindering return-oriented programming using in-place code randomization,'' in
  \emph{2012 IEEE Symposium on Security and Privacy (SP)}.\hskip 1em plus 0.5em
  minus 0.4em\relax IEEE, 2012, pp. 601--615.

\bibitem{pewny2015cross}
J.~Pewny, B.~Garmany, R.~Gawlik, C.~Rossow, and T.~Holz, ``Cross-architecture
  bug search in binary executables,'' in \emph{2015 IEEE Symposium on Security
  and Privacy (SP)}.\hskip 1em plus 0.5em minus 0.4em\relax IEEE, 2015, pp.
  709--724.

\bibitem{prakash2015vfguard}
A.~Prakash, X.~Hu, and H.~Yin, ``vfguard: Strict protection for virtual
  function calls in cots c++ binaries.'' in \emph{NDSS}, 2015.

\bibitem{qiao2017function}
R.~Qiao and R.~Sekar, ``Function interface analysis: A principled approach for
  function recognition in cots binaries,'' in \emph{2017 47th Annual IEEE/IFIP
  International Conference on Dependable Systems and Networks (DSN)}.\hskip 1em
  plus 0.5em minus 0.4em\relax IEEE, 2017, pp. 201--212.

\bibitem{radare2_org}
radare, ``radare2: Unix-like reverse engineering framework and commandline
  tools,'' \url{https://github.com/radare/radare2}, accessed Aug 9, 2019.

\bibitem{shin2015recognizing}
E.~C.~R. Shin, D.~Song, and R.~Moazzezi, ``Recognizing functions in binaries
  with neural networks,'' in \emph{24th USENIX Security Symposium}, 2015, pp.
  611--626.

\bibitem{shoshitaishvili2016sok}
Y.~Shoshitaishvili, R.~Wang, C.~Salls, N.~Stephens, M.~Polino, A.~Dutcher,
  J.~Grosen, S.~Feng, C.~Hauser, C.~Kruegel \emph{et~al.}, ``Sok:(state of) the
  art of war: Offensive techniques in binary analysis,'' in \emph{2016 IEEE
  Symposium on Security and Privacy (SP)}.\hskip 1em plus 0.5em minus
  0.4em\relax IEEE, 2016, pp. 138--157.

\bibitem{skochinsky2012compiler}
I.~Skochinsky, ``Compiler internals: Exceptions and rtti,'' \emph{Recon}, 2012.

\bibitem{van2016tough}
V.~Van Der~Veen, E.~G{\"o}ktas, M.~Contag, A.~Pawoloski, X.~Chen, S.~Rawat,
  H.~Bos, T.~Holz, E.~Athanasopoulos, and C.~Giuffrida, ``A tough call:
  Mitigating advanced code-reuse attacks at the binary level,'' in \emph{2016
  IEEE Symposium on Security and Privacy (SP)}.\hskip 1em plus 0.5em minus
  0.4em\relax IEEE, 2016, pp. 934--953.

\bibitem{wang2015binary}
M.~Wang, H.~Yin, A.~V. Bhaskar, P.~Su, and D.~Feng, ``Binary code continent:
  Finer-grained control flow integrity for stripped binaries,'' in \emph{31st
  Annual Computer Security Applications Conference (ACSAC'15)}.\hskip 1em plus
  0.5em minus 0.4em\relax ACM, 2015, pp. 331--340.

\bibitem{wang2016uroboros}
P.~Wang, Shuai~Wang and D.~Wu, ``Uroboros: Instrumenting stripped binaries with
  static reassembling,'' in \emph{2016 IEEE 23rd International Conference on
  Software Analysis, Evolution, and Reengineering (SANER)}, vol.~1.\hskip 1em
  plus 0.5em minus 0.4em\relax IEEE, 2016, pp. 236--247.

\bibitem{wang2017ramblr}
R.~Wang, Y.~Shoshitaishvili, A.~Bianchi, A.~Machiry, J.~Grosen, P.~Grosen,
  C.~Kruegel, and G.~Vigna, ``Ramblr: Making reassembly great again.'' in
  \emph{NDSS}, 2017.

\bibitem{wang2015reassembleable}
S.~Wang, P.~Wang, and D.~Wu, ``Reassembleable disassembling,'' in \emph{24th
  USENIX Security Symposium}, 2015, pp. 627--642.

\bibitem{wartell2012binary}
R.~Wartell, V.~Mohan, K.~W. Hamlen, and Z.~Lin, ``Binary stirring:
  Self-randomizing instruction addresses of legacy x86 binary code,'' in
  \emph{2012 ACM conference on Computer and communications security}.\hskip 1em
  plus 0.5em minus 0.4em\relax ACM, 2012, pp. 157--168.

\bibitem{williams2016shuffler}
D.~Williams-King, G.~Gobieski, K.~Williams-King, J.~P. Blake, X.~Yuan, P.~Colp,
  M.~Zheng, V.~P. Kemerlis, J.~Yang, and W.~Aiello, ``Shuffler: Fast and
  deployable continuous code re-randomization,'' in \emph{12th USENIX Symposium
  on Operating Systems Design and Implementation (OSDI 16)}, 2016, pp.
  367--382.

\bibitem{williams2020egalito}
D.~Williams-King, H.~Kobayashi, K.~Williams-King, G.~Patterson, F.~Spano, Y.~J.
  Wu, J.~Yang, and V.~P. Kemerlis, ``Egalito: Layout-agnostic binary
  recompilation,'' in \emph{Proceedings of the Twenty-Fifth International
  Conference on Architectural Support for Programming Languages and Operating
  Systems}, 2020, pp. 133--147.

\bibitem{zhang2013practical}
C.~Zhang, T.~Wei, Z.~Chen, L.~Duan, L.~Szekeres, S.~McCamant, D.~Song, and
  W.~Zou, ``Practical control flow integrity and randomization for binary
  executables,'' in \emph{2013 IEEE Symposium on Security and Privacy
  (SP)}.\hskip 1em plus 0.5em minus 0.4em\relax IEEE, 2013, pp. 559--573.

\bibitem{zhang2013control}
M.~Zhang and R.~Sekar, ``Control flow integrity for cots binaries,'' in
  \emph{22nd USENIX Security Symposium}, 2013, pp. 337--352.

\end{thebibliography}
\end{document}